\documentclass[12pt]{article}
\usepackage{amsfonts,graphicx,amssymb,amsmath,amsthm,epsfig}

\allowdisplaybreaks

\begin{document}
\title{\bf Study of Decoupled Cosmological Solutions in $f(\mathbb{R},\mathbb{T})$ Theory}
\author{M. Sharif$^1$ \thanks{msharif.math@pu.edu.pk} and Tayyab Naseer$^{1,2}$ \thanks{tayyabnaseer48@yahoo.com}\\
$^1$ Department of Mathematics and Statistics, The University of Lahore,\\
1-KM Defence Road Lahore, Pakistan.\\
$^2$ Department of Mathematics, University of the Punjab,\\
Quaid-e-Azam Campus, Lahore-54590, Pakistan.}

\date{}
\maketitle
\begin{abstract}
In this paper, we consider a non-static spherical geometry and
formulate its extension for the case of anisotropic matter
configuration through minimal gravitational decoupling in
$f(\mathbb{R},\mathbb{T})$ theory. We apply a particular
transformation only on the radial metric function that divides the
modified field equations into two distinct sectors corresponding to
their parent (original and additional) sources. The unknowns in the
first (isotropic) set are reduced by taking the
Friedmann-Lemaitre-Robertson-Walker cosmic model. We then obtain the
isotropic solution by employing a linear equation of state and
power-law form of the scale factor. The other set involves the
decoupling function and components of an extra source, therefore we
adopt a density-like constraint to close it. Finally, we analyze the
role of this modified gravity and the decoupling parameter on three
different eras of the cosmos by graphically observing the developed
extended solution. It is concluded that the resulting solutions
fulfill all the physical requirements only for the matter and
radiation-dominated eras.
\end{abstract}
{\bf Keywords:} $f(\mathbb{R},\mathbb{T})$ theory; Cosmology;
Gravitational decoupling. \\
{\bf PACS:} 04.50.Kd; 98.80.-k; 04.40.-b.

\section{Introduction}

Various cosmological observations such as Supernova type Ia
\cite{1}, WMAP \cite{1a} and emission of x-rays from galaxies
\cite{1b} confirm rapid expansion of our universe. The plenty of an
unknown force in the cosmos triggers such expansion due to its
immense repulsive effect, named as dark energy. The missing energy
density in the universe can also be compensated by this mysterious
force that prompted astrophysicists to reveal its obscure
characteristics. In this respect, researchers made some geometric
corrections in an Einstein-Hilbert action resulting in different
modified theories of general relativity ($\mathbb{GR}$). The
$f(\mathbb{R})$ theory was an immediate extension obtained by
inserting a generic function of the Ricci scalar in place of
$\mathbb{R}$ in an Einstein-Hilbert action. Various researchers
discussed this theory as the first attempt to study different cosmic
eras such as inflationary as well as present rapid expansion epochs
\cite{2,2a}. Multiple $f(\mathbb{R})$ models have been found to be
stable with the help of different techniques \cite{9,9g}.

The initial idea of the matter-geometry coupling in $f(\mathbb{R})$
theory was presented by Bertolami et al. \cite{10}, who adopted the
Lagrangian as a function of $\mathbb{R}$ and $\pounds_{\mathcal{M}}$
to study the impact of such interaction on heavily objects. This
coupling has recently been generalized at the action level by Harko
et al. \cite{20} by introducing $f(\mathbb{R},\mathbb{T})$ gravity,
in which $\mathbb{T}$ indicates trace of the energy-momentum tensor
($\mathbb{EMT}$). This theory is observed to be non-conserved due to
the inclusion of $\mathbb{T}$ leading to the presence of an
additional force (based on matter variables like density and
pressure \cite{22}) that hampers geodesic motion of the test
particles in their gravitational field. Baffou et al. \cite{22a}
studied the stability of this theory by employing two different
solutions. They also discussed the cosmological dynamics of this
model for low as well as high-redshift regimes, and found their
results consistent with the observational data. Singh and Kumar
\cite{22b} considered the simplest model of the form
$\mathbb{R}+2\varpi\mathbb{T}$ along with bulk viscous fluid and
discussed the expansion of our cosmos for certain positive and
negative values of the parameter $\varpi$. Sharif and Zubair
\cite{22c} explored that this theory can reproduce any cosmological
evolution like phantom/non-phantom regimes and $\Lambda$ cold dark
matter, etc. We have discussed anisotropic spherical structures in
this context and found physically feasible models for certain values
of the model parameter \cite{22d}.

The cosmos is isotropic and homogeneous at scales larger than
$300h^{-1}$ Mpc according to the cosmological principle \cite{9a}.
In agreement with this principle, researchers often employ
Friedmann-Lemaitre-Robertson-Walker (FLRW) metric to discuss
expanding behavior of the universe. Different cosmological surveys
in the last two decades indicate that this cosmos deviates from
being isotropic. For instance, the exploration of inhomogeneous
Supernova Ia noticed small discrepancy in an isotropic nature
\cite{10aaa}. Some other plausible inconsistencies have been
determined with the help of multiple tools such as radio sources
\cite{10aa}, infrared galaxies \cite{10ab} and gamma-ray bursts
\cite{10ac}, etc. Migkas and Reiprich \cite{10a} recently examined
the isotropy of this universe with the help of the
direction-dependent behavior of x-rays originating from galactic
structures. Later on, the same technique has been implemented on
some other clusters of galaxies from which the cosmos was discovered
to be anisotropic \cite{10b}. Therefore, it is highly important to
construct anisotropic solutions that would be helpful in
understanding the cosmic origin as well as its ultimate fate
properly.

It is a difficult task to determine solutions of the field equations
corresponding to any self-gravitating body due to the entanglement
of highly non-linear terms. There are multiple approaches introduced
in the literature and the most recent technique in this regard is
the gravitational decoupling that extends isotropic structures to
anisotropic domains. This strategy provides transformations on
metric functions that decouple the field equations into two distinct
sectors, each of which corresponds to its original source. Both sets
are solved independently and then the superposition principle is
applied to get the solution of the total setup. Ovalle \cite{11}
pioneered this technique and obtained exact anisotropic models
corresponding to spherical geometry in the braneworld through
minimal geometric deformation (MGD) that allows only $g_{rr}$
component to transform. Later, Ovalle et al. \cite{13} extended the
isotropic solution to two different anisotropic solutions in
$\mathbb{GR}$ and checked the role of the decoupling parameter on
anisotropy. Gabbanelli et al. \cite{14} calculated extended versions
of the Durgapal-Fuloria isotropic geometry through this technique.

Multiple anisotropic extensions of different isotropic models such
as Krori-Barua \cite{15}, Heintzmann \cite{17} and Tolman VII
\cite{19} have been developed with the help of MGD strategy.
Cede{\~n}o and Contreas \cite{19a} employed this scheme on
Kantowski-Sachs and FLRW geometries to find their respective new
cosmological solutions representing anisotropic spacetimes. Sharif
and Majid \cite{16a} formulated new cosmic model by extending
isotropic FLRW spacetime through MGD and explored physical
feasibility of different eras of our universe in Brans-Dicke theory.
The isotropic Korkina-Orlyanskii \cite{16b} as well as Tolman VII
ansatz \cite{16c} have also been extended to multiple viable and
stable anisotropic solutions in $f(\mathbb{R},\mathbb{T})$ gravity.
Sharif and his collaborators established several acceptable
anisotropic versions of different isotropic models through this
technique in different modified theories such as
$f(\mathbb{R}),~f(\mathbb{G})$ and Brans-Dicke scenario \cite{16}.
We have discussed various anisotropic compact stars and explored
their stability conditions in a matter-geometry coupled gravity
\cite{21,21a}.

This article aims to extend isotropic FLRW solution to anisotropic
model with the help of MGD scheme in the context of
$f(\mathbb{R},\mathbb{T})$ theory. The paper has the following
pattern. The basic definitions of this modified gravity and the
insertion of new gravitating anisotropic source are studied in
section \textbf{2}. In section \textbf{3}, we decouple the field
equations through a particular transformation and deduce two
different sets that belong to their original sources. The graphical
behavior of state determinants, viability and stability criteria are
analyzed in section \textbf{4}. Section \textbf{5} concludes all our
results.

\section{$f(\mathbb{R},\mathbb{T})$ Formalism}

The $f(\mathbb{R},\mathbb{T})$ field equations can be obtained by
modifying the Einstein-Hilbert action in the following form
\begin{equation}\label{1}
\mathcal{A}_{f(\mathbb{R},\mathbb{T})}=\int\sqrt{-g}\bigg\{\frac{f(\mathbb{R},\mathbb{T})}{16\pi}+\pounds_{\mathcal{M}}
+\delta\pounds_{\mathcal{Y}}\bigg\}d^{4}x,
\end{equation}
where $\pounds_{\mathcal{M}}$ and $\pounds_{\mathcal{Y}}$ are
Lagrangian densities of the matter configuration (or seed source)
and an additional source, respectively. Also, $g$ indicates
determinant of the metric tensor ($g_{\lambda\zeta}$) and $\delta$
symbolizes the decoupling parameter. The variational principle can
be implemented on the action \eqref{1} that provides the following
modified field equations
\begin{equation}\label{2}
\mathbb{G}_{\lambda\zeta}=8\pi \mathbb{T}_{\lambda\zeta}^{(tot)},
\end{equation}
where an Einstein tensor $\mathbb{G}_{\lambda\zeta}$ is further
expressed in terms of geometric quantities as
$\mathbb{G}_{\lambda\zeta}=\mathbb{R}_{\lambda\zeta}-\frac{1}{2}\mathbb{R}g_{\lambda\zeta}$,
whereas the total matter distribution on the right hand side is
classified as
\begin{equation}\label{3}
\mathbb{T}_{\lambda\zeta}^{(tot)}=\mathbb{T}_{\lambda\zeta}^{(eff)}+\delta
\mathcal{Y}_{\lambda\zeta}=\frac{1}{f_{\mathbb{R}}}\mathbb{T}_{\lambda\zeta}+\mathbb{T}_{\lambda\zeta}^{(\mathbf{D})}+\delta
\mathcal{Y}_{\lambda\zeta}.
\end{equation}
The additional source ($\mathcal{Y}_{\lambda\zeta}$) in this case
produces anisotropy in the considered setup whose influence can be
controlled by the parameter $\delta$. The quantity
$\mathbb{T}_{\lambda\zeta}^{(\mathbf{D})}$ exhibits the contribution
of modified gravity and is presented as
\begin{eqnarray}
\nonumber \mathbb{T}_{\lambda\zeta}^{(\mathbf{D})}&=&\frac{1}{8\pi
f_{\mathbb{R}}}
\bigg[f_{\mathbb{T}}\mathbb{T}_{\lambda\zeta}+\bigg\{\frac{\mathbb{R}}{2}\bigg(\frac{f}{\mathbb{R}}-f_{\mathbb{R}}\bigg)
-\pounds_{\mathcal{M}}f_{\mathbb{T}}\bigg\}g_{\lambda\zeta}\\\label{4}
&-&(g_{\lambda\zeta}\Box-\nabla_{\lambda}\nabla_{\zeta})f_{\mathbb{R}}+2f_{\mathbb{T}}g^{\eta\beta}\frac{\partial^2
\pounds_{\mathcal{M}}}{\partial g^{\lambda\zeta}\partial
g^{\eta\beta}}\bigg],
\end{eqnarray}
where $f_{\mathbb{R}}=\frac{\partial
f(\mathbb{R},\mathbb{T})}{\partial \mathbb{R}}$ and
$f_{\mathbb{T}}=\frac{\partial f(\mathbb{R},\mathbb{T})}{\partial
\mathbb{T}}$. Moreover, $\nabla_\zeta$ and $\Box\equiv
\frac{1}{\sqrt{-g}}\partial_\lambda\big(\sqrt{-g}g^{\lambda\zeta}\partial_{\zeta}\big)$
denote the covariant derivative and the D'Alembert operator,
respectively. The field equations in this scenario has the following
trace
\begin{align}\nonumber
&3\nabla^{\zeta}\nabla_{\zeta}f_\mathbb{R}+\mathbb{R}f_\mathbb{R}-\mathbb{T}(f_\mathbb{T}+1)-2f+4f_\mathbb{T}\pounds_\mathcal{M}
-2f_\mathbb{T}g^{\eta\beta}g^{\lambda\zeta}\frac{\partial^2\pounds_\mathcal{M}}{\partial
g^{\eta\beta}\partial g^{\lambda\zeta}}=0.
\end{align}
The nature of seed matter source is considered to be perfect, thus
the corresponding $\mathbb{EMT}$ is given as
\begin{equation}\label{5}
\mathbb{T}_{\lambda\zeta}=(\mu+P)\mathrm{K}_{\lambda}\mathrm{K}_{\zeta}+Pg_{\lambda\zeta},
\end{equation}
where $\mathrm{K}_{\zeta},~P$ and $\mu$ are the four-velocity,
pressure and energy density, respectively.

We consider a non-static spherical geometry describing the interior
region by the line element as
\begin{equation}\label{6}
ds^{2}=-e^{\alpha_{1}}dt^{2}+e^{\alpha_{2}}dr^{2}+\mathcal{C}^{2}(d\theta^{2}+{\sin^{2}\theta}{d\phi^2}),
\end{equation}
where $\alpha_{1}=\alpha_{1}(t,r),~\alpha_{2}=\alpha_{2}(t,r)$ and
$\mathcal{C}=\mathcal{C}(t,r)$. The four-velocity is defined as
\begin{equation}\label{7}
\mathrm{K}^\zeta=(e^{\frac{-\alpha_{1}}{2}},0,0,0),
\end{equation}
with $\mathrm{K}_{\zeta}\mathrm{K}^{\zeta}=-1$. We shall require to
take a particular $f(\mathbb{R},\mathbb{T})$ model to discuss
physical acceptability of the cosmological solutions. The most
effective model in this regard that completely deforms the modified
field equations is given as
\begin{equation}\label{7a}
f(\mathbb{R},\mathbb{T})=f_1(\mathbb{R})+
f_2(\mathbb{T})=\mathbb{R}+2\varpi\mathbb{T},
\end{equation}
where a real-valued coupling parameter $\varpi$ helps to determine
the role of this theory on geometrical structures and
$\mathbb{T}=-\mu+3P$. Houndjo and Piattella \cite{38} found that the
attributes of holographic dark energy may be regenerated by studying
the pressureless matter source along with this model. A lot of
successful analysis has been done by researchers with the help of
this functional form \cite{39}. The field equations \eqref{2} take
the form for the metric \eqref{6} and functional \eqref{7a} as
\begin{align}\nonumber
8\pi\left(\mu-\delta\mathcal{Y}_{0}^{0}\right)+\varpi\left(3\mu-P\right)
&=\frac{1}{\mathcal{C}^2}-\frac{e^{-\alpha_{2}}}{\mathcal{C}}\bigg(\frac{\mathcal{C}^{'2}}{\mathcal{C}}
-\mathcal{C}^{'}\alpha_{2}^{'}+2\mathcal{C}^{''}\bigg)\\\label{8}
&+\frac{e^{-\alpha_{1}}}{\mathcal{C}}\bigg(\frac{\dot{\mathcal{C}}^2}{\mathcal{C}}
+\dot{\mathcal{C}}\dot{\alpha_{2}}\bigg),\\\nonumber
8\pi\left(P+\delta\mathcal{Y}_{1}^{1}\right)-\varpi\left(\mu-3P\right)
&=-\frac{1}{\mathcal{C}^2}-\frac{e^{-\alpha_{1}}}{\mathcal{C}}\bigg(\frac{\dot{\mathcal{C}}^2}{\mathcal{C}}
-\dot{\mathcal{C}}\dot{\alpha_{1}}+2\ddot{\mathcal{C}}\bigg)\\\label{9}
&+\frac{e^{-\alpha_{2}}}{\mathcal{C}}\bigg(\frac{\mathcal{C}^{'2}}{\mathcal{C}}+\mathcal{C}^{'}\alpha_{1}^{'}\bigg),\\\nonumber
8\pi\left(P+\delta\mathcal{Y}_{2}^{2}\right)-\varpi\left(\mu-3P\right)
&=e^{-\alpha_{2}}\bigg(\frac{\alpha_{1}^{'2}}{4}-\frac{\alpha_{1}^{'}\alpha_{2}
^{'}}{4}+\frac{\alpha_{1}^{''}}{2}+\frac{\mathcal{C}^{''}}{\mathcal{C}}\bigg)\\\nonumber
&-e^{-\alpha_{1}}\bigg(\frac{\dot{\alpha_{2}}^{2}}{4}-\frac{\dot{\alpha_{1}}\dot{\alpha_{2}}}{4}+\frac{\ddot{\alpha_{2}}}{2}
+\frac{\ddot{\mathcal{C}}}{\mathcal{C}}\bigg)\\\label{9a}
&+\frac{1}{2\mathcal{C}}\big\{e^{-\alpha_{2}}\mathcal{C}^{'}\big(\alpha_{1}^{'}
-\alpha_{2}^{'}\big)+e^{-\alpha_{1}}\dot{\mathcal{C}}\big(\dot{\alpha_{1}}-\dot{\alpha_{2}}\big)\big\},\\\label{9b}
8\pi\delta\mathcal{Y}^{0}_{1}&=\frac{e^{-\alpha_{1}}}{\mathcal{C}}\big(2\dot{\mathcal{C}'}-\dot{\mathcal{C}}\alpha_{1}^{'}
-\mathcal{C}^{'}\dot{\alpha_{2}}\big),
\end{align}
where $.=\frac{\partial}{\partial{t}}$ and
$'=\frac{\partial}{\partial{r}}$. Moreover, the expressions
multiplied by $\varpi$ on the right side appear due to the modified
gravity. Equations \eqref{8}-\eqref{9b} represent anisotropic fluid
distribution when $\mathcal{Y}_{1}^{1} \neq \mathcal{Y}_{2}^{2}$.
The coupling between geometry and matter in modified theories yields
non-conserved nature of the $\mathbb{EMT}$, opposing $\mathbb{GR}$
and $f(\mathbb{R})$ gravity. Hence, the non-conservation in this
framework can be seen as
\begin{equation}\label{g11}
\nabla^\lambda\mathbb{T}_{\lambda\zeta}=\frac{f_\mathbb{T}}{8\pi-f_\mathbb{T}}\bigg[(\mathbb{T}_{\lambda\zeta}
+\Theta_{\lambda\zeta})\nabla^\lambda\ln{f_\mathbb{T}}+\nabla^\lambda\Theta_{\lambda\zeta}
-\frac{8\pi\delta}{f_\mathbb{T}}\nabla^\lambda\mathcal{Y}_{\lambda\zeta}
-\frac{1}{2}g_{\eta\beta}\nabla_\zeta\mathbb{T}^{\eta\beta}\bigg],
\end{equation}
where
$\Theta_{\lambda\zeta}=g_{\lambda\zeta}\pounds_\mathcal{M}-2\mathbb{T}_{\lambda\zeta}-2g^{\eta\beta}\frac{\partial^2
\pounds_\mathcal{M}}{\partial g^{\lambda\zeta}\partial
g^{\eta\beta}}$. In this case, we consider $\pounds_\mathcal{M}=P$
which leads to $\frac{\partial^2 \pounds_\mathcal{M}}{\partial
g^{\lambda\zeta}\partial g^{\eta\beta}}=0$.

\section{Gravitational Decoupling}

In this section, we reduce degrees of freedom of set of highly
non-linear differential equations \eqref{8}-\eqref{9b} by employing
MGD approach. This set contains nine unknowns
$(\alpha_{1},\alpha_{2},\mathcal{C},\mu,P,\mathcal{Y}^{0}_{0},\mathcal{Y}^{1}_{1},\mathcal{Y}^{2}_{2},\mathcal{Y}^{0}_{1})$,
indicating the under-determined system. We are therefore required to
apply some constraints so that the system can be solved easily. As
we have added an additional anisotropic source in the original
matter source, the MGD scheme is now used to determine two
independent sets of field equations which will be solved separately.
Initially, Ovalle \cite{11} introduced the transformation on the
radial metric component to decouple the field equations
corresponding to static source whose extended form for non-static
scenario is taken as
\begin{equation}\label{16}
e^{-\alpha_{2}(t,r)} \mapsto e^{-\alpha_{3}(t,r)}\{1+\delta
\bar{h}(t,r)\}.
\end{equation}
After applying the transformation \eqref{16} on
Eqs.\eqref{8}-\eqref{9b}, the seed isotropic source can be recovered
for $\delta=0$ as
\begin{align}\nonumber
8\pi\mu&=\varpi{P}-3\varpi\mu+\frac{\dot{\mathcal{C}}e^{-\alpha_{1}}\dot{\alpha_{3}}}{\mathcal{C}}+\frac{\dot{\mathcal{C}}^2
e^{-\alpha_{1}}}{\mathcal{C}^2}+\frac{\mathcal{C}'e^{-\alpha_{3}}\alpha_{3}'}{\mathcal{C}}+\frac{1}{\mathcal{C}^2}\\\label{17}
&-\frac{\mathcal{C}'^2e^{-\alpha_{3}}}{\mathcal{C}^2}-\frac{2\mathcal{C}''e^{-\alpha_{3}}}{\mathcal{C}},\\\nonumber
8\pi{P}&=\varpi\mu-3\varpi{P}+\frac{\mathcal{C}'\alpha_{1}'e^{-\alpha_{3}}}{\mathcal{C}}+\frac{\dot{\mathcal{C}}
e^{-\alpha_{1}}\dot{\alpha_{1}}}{\mathcal{C}}-\frac{\dot{\mathcal{C}}^2e^{-\alpha_{1}}}{\mathcal{C}^2}
-\frac{1}{\mathcal{C}^2}\\\label{17a}
&-\frac{2\ddot{\mathcal{C}}e^{-\alpha_{1}}}{\mathcal{C}}+\frac{\mathcal{C}'^2e^{-\alpha_{3}}}{\mathcal{C}^2},\\\nonumber
8\pi{P}&=\varpi\mu-3\varpi{P}+\frac{\mathcal{C}'\alpha_{1}'e^{-\alpha_{3}}}{2\mathcal{C}}+\frac{\dot{\mathcal{C}}
e^{-\alpha_{1}}\dot{\alpha_{1}}}{2\mathcal{C}}-\frac{\dot{\mathcal{C}}e^{-\alpha_{1}}\dot{\alpha_{3}}}{2\mathcal{C}}
-\frac{\ddot{\mathcal{C}}e^{-\alpha_{1}}}{\mathcal{C}}\\\nonumber
&-\frac{\mathcal{C}'e^{-\alpha_{3}}\alpha_{3}'}{2\mathcal{C}}+\frac{\mathcal{C}''e^{-\alpha_{3}}}{\mathcal{C}}
-\frac{1}{4}\alpha_{1}'e^{-\alpha_{3}}\alpha_{3}'+\frac{1}{4}\alpha_{1}'^2e^{-\alpha_{3}}
+\frac{1}{2}\alpha_{1}''e^{-\alpha_{3}}\\\label{17b}
&+\frac{1}{4}e^{-\alpha_{1}}\dot{\alpha_{1}}\dot{\alpha_{3}}-\frac{1}{4}e^{-\alpha_{1}}\dot{\alpha_{3}}^2
-\frac{1}{2}e^{-\alpha_{1}}\ddot{\alpha_{3}},\\\label{17c}
0&=-\frac{\dot{\mathcal{C}}e^{-\alpha_{1}}\alpha_{1}'}{\mathcal{C}}-\frac{\mathcal{C}'e^{-\alpha_{1}}\dot{\alpha_{3}}}{\mathcal{C}}
+\frac{2\dot{\mathcal{C}'}e^{-\alpha_{1}}}{\mathcal{C}}.
\end{align}

We can observe that we still require an extra constraint to solve
the system \eqref{17}-\eqref{17c} analytically. Here, an important
equation of state $\mathbb{E}o\mathbb{S}$ that establishes the
relation between energy density and pressure is taken as
\begin{equation}\label{16a}
P=\omega\mu,
\end{equation}
where $\omega$ is the parameter whose different values are used in
the literature to discuss different eras of our universe. For
instance
\begin{itemize}
\item $\omega=\frac{1}{3}$ \quad $\Rightarrow$ \quad radiation-dominated era,
\item $\omega=0$ \quad $\Rightarrow$ \quad matter-dominated era (or dust era as $P=0$),
\item $\omega=-1$ \quad $\Rightarrow$ \quad vacuum energy dominated era.
\end{itemize}
Equations \eqref{17} and \eqref{17a} provide explicit forms of the
energy density and pressure after using $\mathbb{E}o\mathbb{S}$
\eqref{16a} as
\begin{align}\nonumber
\mu&= \frac{ e^{-\alpha_{1}-\alpha_{3}}}{(3 \varpi-\omega\varpi +8
\pi ) \mathcal{C}^2}\big(C' C e^{\alpha_{1}} \alpha_{3}'-2 C'' C
e^{\alpha_{1}}-C'^2 e^{\alpha_{1}}\\\label{16b}&+\dot{C} C
e^{\alpha_{3}} \dot{\alpha_{3}}+\dot{C}^2
e^{\alpha_{3}}+e^{\alpha_{1}+\alpha_{3}}\big),\\\nonumber
P&=\frac{\omega e^{-\alpha_{1}-\alpha_{3}}}{(3 \omega\varpi -\varpi
+8 \pi \omega ) \mathcal{C}^2} \big(\dot{C} C \dot{\alpha_{1}}
e^{\alpha_{3}}+C' C e^{\alpha_{1}} \alpha_{1}'+C'^2
e^{\alpha_{1}}\\\label{16c}&-2 \ddot{C} C e^{\alpha_{3}}-\dot{C}^2
e^{\alpha_{3}}-e^{\alpha_{1}+\alpha_{3}}\big).
\end{align}

Furthermore, the additional source can be represented by the
following set of equations as
\begin{align}\label{18}
8\pi\mathcal{Y}^{0}_{0}&=\frac{\dot{\mathcal{C}} \dot{\bar{h}}
e^{-\alpha_{1}}}{\mathcal{C} (\delta  \bar{h}+1)}+\frac{\mathcal{C}'
\bar{h}' e^{-\alpha_{3}}}{\mathcal{C}}-\frac{\mathcal{C}' \bar{h}
e^{-\alpha_{3}} \alpha_{3}'}{\mathcal{C}}+\frac{\mathcal{C}'^2
\bar{h} e^{-\alpha_{3}}}{\mathcal{C}^2}+\frac{2 \mathcal{C}''
\bar{h} e^{-\alpha_{3}}}{\mathcal{C}},\\\label{18a}
8\pi\mathcal{Y}^{1}_{1}&=\frac{\mathcal{C}' \bar{h} \alpha_{1}'
e^{-\alpha_{3}}}{\mathcal{C}}+\frac{\mathcal{C}'^2
\bar{h}e^{-\alpha_{3}}}{\mathcal{C}^2},\\\nonumber
8\pi\mathcal{Y}^{2}_{2}&=\frac{\dot{\mathcal{C}} \dot{\bar{h}}
e^{-\alpha_{1}}}{2\mathcal{C} (\delta \bar{h}+1)}+\frac{\mathcal{C}'
\bar{h}' e^{-\alpha_{3}}}{2 \mathcal{C}}+\frac{\mathcal{C}' \bar{h}
\alpha_{1}' e^{-\alpha_{3}}}{2 \mathcal{C}}-\frac{\mathcal{C}'
\bar{h} e^{-\alpha_{3}} \alpha_{3}'}{2
\mathcal{C}}+\frac{\mathcal{C}'' \bar{h}
e^{-\alpha_{3}}}{\mathcal{C}}\\\nonumber&-\frac{\dot{\bar{h}}
e^{-\alpha_{1}} \dot{\alpha_{1}}}{4 (\delta
\bar{h}+1)}+\frac{\dot{\bar{h}} e^{-\alpha_{1}} \dot{\alpha_{3}}}{2
(\delta \bar{h}+1)}-\frac{3 \delta \dot{\bar{h}}^2
e^{-\alpha_{1}}}{4 (\delta \bar{h}+1)^2}+\frac{\ddot{\bar{h}}
e^{-\alpha_{1}}}{2 (\delta \bar{h}+1)}+\frac{1}{4} \bar{h}'
\alpha_{1}' e^{-\alpha_{3}}\\\label{18b}&-\frac{1}{4} \bar{h}
\alpha_{1}' e^{-\alpha_{3}} \alpha_{3}'+\frac{1}{4} \bar{h}
\alpha_{1}'^2 e^{-\alpha_{3}}+\frac{1}{2} \bar{h} \alpha_{1}''
e^{-\alpha_{3}},\\\label{18c}
8\pi\mathcal{Y}^{0}_{1}&=\frac{\mathcal{C}' \dot{\bar{h}}
e^{-\alpha_{1}}}{\mathcal{C} (\delta \bar{h}+1)}.
\end{align}

\section{Anisotropic Cosmological Solution}

In this section, we assume a seed isotropic matter source to be the
FLRW metric representing homogeneous universe in order to formulate
anisotropic cosmological models as
\begin{equation}\label{32}
ds^{2}=-dt^{2}+\mathrm{a}^2(t)\bigg(\frac{dr^{2}}{1-\mathrm{k}r^2}+r^2d\theta^{2}+r^2{\sin^{2}\theta}{d\phi^2}\bigg),
\end{equation}
where $\mathrm{a}(t)$ refers to the scale factor. As our universe
expands, the distance between distinct points changes which can be
measured by $\mathrm{a}(t)$. Also, $\mathrm{k}$ indicates the
curvature parameter whose different values ($1,0,-1$) correspond to
close, flat and open cosmic models, respectively. The line elements
\eqref{6} and \eqref{32} provide the following relations between
their metric coefficients as
\begin{align}\label{33}
e^{\alpha_{1}(t,r)}=1, \quad
e^{\alpha_{3}(t,r)}=\frac{\mathrm{a}^2(t)}{1-\mathrm{k}r^2}, \quad
\mathcal{C}(t,r)=r \mathrm{a}(t).
\end{align}
After using these relations, Eqs.\eqref{16b} and \eqref{16c} reduce
to
\begin{align}\label{36}
\mu&=\frac{3\left(\dot{\mathrm{a}}^2+\mathrm{k}\right)}{\mathrm{a}^2
\{8 \pi -\varpi (\omega -3)\}}, \\\label{37} P&=-\frac{\omega
\left(2\mathrm{a}\ddot{\mathrm{a}}+\dot{\mathrm{a}}^2+\mathrm{k}\right)}{\mathrm{a}^2
\{\varpi (3 \omega -1)+8 \pi \omega \}}.
\end{align}
The above two equations contain three unknowns, thus we take the
scale factor in its power-law functional form \cite{8a} to fully
determine the solution corresponding to isotropic source as
\begin{equation}\label{37a}
\mathrm{a}(t)=\mathrm{a}_{o}t^\eta
\end{equation}
where $\mathrm{a}_{o}$ and $\eta$ are the present value of the scale
factor and a positive constant, respectively. Moreover, the system
representing the anisotropic set can be rewritten in the form of
FLRW spacetime
\begin{align}\label{38}
8\pi\mathcal{Y}^{0}_{0}&=\frac{1}{\mathrm{a}^2}\bigg[\frac{\mathrm{a}
\dot{\mathrm{a}} \dot{\bar{h}}}{\delta \bar{h}+1}-\bigg(\mathrm{k}
r-\frac{1}{r}\bigg) \bar{h}'-\bigg(3 \mathrm{k}-\frac{1}{r^2}\bigg)
\bar{h}\bigg],
\\\label{38a}8\pi\mathcal{Y}^{1}_{1}&=\frac{\left(1-\mathrm{k} r^2\right) \bar{h}}{r^2
\mathrm{a}^2},\\\label{38b}8\pi\mathcal{Y}^{2}_{2}&=\frac{1}{4}\bigg[\frac{6
\dot{\mathrm{a}} \dot{\bar{h}}}{\mathrm{a}(\delta
\bar{h}+1)}-\frac{2 \left(\mathrm{k} r^2-1\right) \bar{h}'+4
\mathrm{k} r \bar{h}}{r \mathrm{a}^2}+\frac{2 \ddot{\bar{h}} (\delta
\bar{h}+1)-3 \delta \dot{\bar{h}}^2}{(\delta
\bar{h}+1)^2}\bigg],\\\label{38c}8\pi\mathcal{Y}^{0}_{1}&=\frac{\dot{\bar{h}}}{\delta
r \bar{h}+r}.
\end{align}

The extension of the existing isotropic solution delineating the
FLRW metric can be obtained by taking the linear combination of both
the solutions corresponding to their parent sources as
\begin{align}\label{39}
\hat{\mu}&=\mu-\delta\mathcal{Y}^{0}_{0},\\\label{39a}
\hat{P}_r&=P+\delta\mathcal{Y}^{1}_{1},\\\label{39b}
\hat{P}_\bot&=P+\delta\mathcal{Y}^{2}_{2},\\\label{39c}0&=\delta\mathcal{Y}^{0}_{1},
\end{align}
where the anisotropic factor is defined as
$\hat{\Delta}=\hat{P}_\bot-\hat{P}_r=\delta(\mathcal{Y}^{2}_{2}-\mathcal{Y}^{1}_{1})$,
indicating that $\delta=0$ would lead to vanishing anisotropy. We
now have five unknowns
($\mathcal{Y}^{0}_{0},\mathcal{Y}^{1}_{1},\mathcal{Y}^{2}_{2},\mathcal{Y}^{0}_{1},\bar{h}$)
in the system \eqref{38}-\eqref{38c}, therefore a constraint on
additional source is required to determine the exact solution. In
view of this, the most effective (density-like) constraint is taken
as
\begin{equation}\label{40}
\mu=\mathcal{Y}^{0}_{0}.
\end{equation}
We observe from Eq.\eqref{39c} that $\delta\neq0$ results in
$\mathcal{Y}^{0}_{1}=0$ that produces $\dot{\bar{h}}=0$ after
combining with Eq.\eqref{38c}. Using this together with
Eqs.\eqref{36} and \eqref{38} in \eqref{40}, we have the deformation
function as
\begin{equation}\label{41}
\bar{h}=\frac{r^2(\dot{\mathrm{a}}^2+\mathrm{k})}{(1-\mathrm{k}r^2)\{8\pi-\varpi(\omega-3)\}}
+\frac{\mathbb{C}_1}{r(1-\mathrm{k}r^2)},
\end{equation}
where $\mathbb{C}_1$ appears as an integration constant. Finally, we
have the anisotropic solution in the following
\begin{align}\nonumber
\hat{\mu}&=\frac{1}{8\mathrm{a}^2\{8\pi-\varpi(\omega-3)\}}\big[24(\dot{\mathrm{a}}^2+\mathrm{k})+\delta
\big\{\pi\big(-\delta{r^3}\dot{\mathrm{a}}^2+\varpi\delta\mathbb{C}_1\omega-3\varpi\delta\mathbb{C}_1\\\nonumber
&+8\pi\big(-\delta\mathbb{C}_1+\mathrm{k}r^3-r\big)-\varpi\mathrm{k}r^3\omega+3\varpi\mathrm{k}r^3
-\delta\mathrm{k}r^3+\varpi{r}\omega-3\varpi{r}\big)\big\}^{-1}\\\nonumber
&\times\big\{3 \delta  r^3 \dot{\mathrm{a}}^4+\dot{\mathrm{a}}^2
\big(2 r^3 \mathrm{a} (8 \pi -\varpi (\omega -3))
\ddot{\mathrm{a}}+3 \big(8 \pi \big(\delta \mathbb{C}_1-\mathrm{k}
r^3+r\big)-\varpi \delta\\\nonumber &\times \mathbb{C}_1 (\omega
-3)+\mathrm{k} r^3 (\varpi  (\omega -3)+2 \delta )-\varpi  r (\omega
-3)\big)\big)-3 \mathrm{k} \big(\varpi \delta \mathbb{C}_1 (\omega
-3)\\\label{42} &-8 \pi \big(\delta \mathbb{C}_1-\mathrm{k}
r^3+r\big)-\mathrm{k} r^3 (\varpi  (\omega -3)+\delta )+\varpi  r
(\omega -3)\big)\big\}\big],\\\nonumber
\hat{P}_r&=\frac{1}{8\mathrm{a}^2\{8\pi\omega+\varpi(3\omega-1)\}\{\pi{r^3}(8\pi-\varpi(\omega-3))\}}
\big[\delta  \{\varpi  (3 \omega -1)+8 \pi  \omega\} \\\nonumber
&\times\big\{r^3 \dot{\mathrm{a}}^2+\mathbb{C}_1 (-\varpi\omega+3
\varpi +8 \pi )+\mathrm{k} r^3\big\}-8 \pi r^3 \omega \{8 \pi
-\varpi (\omega -3)\}\\\label{42a}
&\times\big(2\mathrm{a}\ddot{\mathrm{a}}+\dot{\mathrm{a}}^2+\mathrm{k}\big)\big],\\\nonumber
\hat{P}_\bot&=-\frac{\omega\left(2\mathrm{a}\ddot{\mathrm{a}}+\dot{\mathrm{a}}^2+\mathrm{k}\right)}{\mathrm{a}^2
\{\varpi(3\omega-1)+8\pi\omega\}}+\frac{\delta}{32\pi}\bigg[\frac{2\{2\mathrm{k}r^3+\mathbb{C}_1(\varpi(\omega-3)-8\pi)+
2r^3\dot{\mathrm{a}}^2\}}{r^3\mathrm{a}^2\{8\pi-\varpi(\omega-3)\}}\\\nonumber
&-\frac{12r^3\dot{\mathrm{a}}^2\ddot{\mathrm{a}}}{\mathrm{a}}\big\{\varpi
\delta  \mathbb{C}_1 (\omega -3)-\delta r^3 \dot{\mathrm{a}}^2+8 \pi
(-\delta \mathbb{C}_1+\mathrm{k} r^3-r)-\mathrm{k} r^3 (\varpi
(\omega -3)\\\nonumber &+\delta)+\varpi  r (\omega
-3)\big\}^{-1}+4r^3\big\{\varpi \delta  \mathbb{C}_1 (\omega
-3)-\delta r^3 \dot{\mathrm{a}}^2+8 \pi (-\delta
\mathbb{C}_1+\mathrm{k} r^3-r)\\\nonumber &-\mathrm{k} r^3 (\varpi
(\omega -3)+\delta)+\varpi  r (\omega
-3)\big\}^{-2}\big\{\dddot{\mathrm{a}} \dot{\mathrm{a}} \big(\delta
r^3 \dot{a}^2+8 \pi \big(\delta \mathbb{C}_1-\mathrm{k}
r^3+r\big)\\\nonumber &-\varpi \delta \mathbb{C}_1 (\omega
-3)+\mathrm{k} r^3 (\varpi (\omega -3)+\delta )-\varpi  r (\omega
-3)\big)+\ddot{\mathrm{a}}^2 \big(\mathrm{k} r^3 (\varpi  (\omega
-3)\\\label{42b} &+\delta )-2 \delta r^3 \dot{\mathrm{a}}^2-\varpi
\delta  \mathbb{C}_1 (\omega -3)+8 \pi \big(\delta
\mathbb{C}_1-\mathrm{k} r^3+r\big)-\varpi r (\omega
-3)\big)\big\}\bigg],\\\nonumber
\hat{\Delta}&=-\frac{\delta}{16\pi{r^3}}\bigg[\frac{3\mathbb{C}_1}{\mathrm{a}^2}+\frac{6r^6\dot{\mathrm{a}}^2\ddot{\mathrm{a}}}
{\mathrm{a}}\big\{\varpi\delta\mathbb{C}_1(\omega-3)-\delta{r^3}\dot{\mathrm{a}}^2+8\pi\big(\mathrm{k}r^3
-\delta\mathbb{C}_1-r\big)\\\nonumber
&-\mathrm{k}r^3(\varpi(\omega-3)+\delta)+\varpi{r}(\omega-3)\big\}^{-1}-2r^6\big\{\varpi
\delta  \mathbb{C}_1 (\omega -3)-\delta r^3 \dot{\mathrm{a}}^2+8
\pi\\\nonumber &\times(\mathrm{k} r^3-\delta
\mathbb{C}_1-r)-\mathrm{k} r^3 (\varpi (\omega -3)+\delta)+\varpi r
(\omega
-3)\big\}^{-2}\big\{\dddot{\mathrm{a}}\dot{\mathrm{a}}\big(\delta{r^3}\dot{\mathrm{a}}^2\\\nonumber
&-\varpi
\delta\mathbb{C}_1(\omega-3)+8\pi\big(\delta\mathbb{C}_1-\mathrm{k}r^3+r\big)+\mathrm{k}
r^3(\varpi(\omega-3)+\delta)-\varpi{r}(\omega-3)\big)\\\nonumber
&+\ddot{\mathrm{a}}^2
\big(8\pi\big(\delta\mathbb{C}_1-\mathrm{k}r^3+r\big)-2\delta{r^3}\dot{\mathrm{a}}^2-\varpi\delta\mathbb{C}_1
(\omega-3)+\mathrm{k} r^3(\varpi(\omega-3)+\delta)\\\label{42c}
&-\varpi{r}(\omega-3)\big)\big\}\bigg].
\end{align}

We now investigate some salient features of the anisotropic FLRW
model corresponding to three different values of $\omega$ in the
following subsections. The deceleration parameter should lie within
$[-1,0]$ according to recent experiments implying $\eta>1$
\cite{24ab}, thus we take it $1.1$ along with $\mathbb{C}_1=0.001$,
$\mathrm{k}=0$. Moreover, we choose $\varpi=0.3$ and
$\delta=0.1,0.9$ to analyze the impact of decoupling parameter on
different models. The $\mathbb{EMT}$ is a source that describes the
inner fluid distribution of a celestial system on which some
constraints are imposed to check whether normal matter exists or
not, known as energy conditions $(\mathbb{EC}s)$. To ensure the
validity of our developed solution \eqref{42}-\eqref{42c}, we
explore these bounds graphically. They are classified in
$f(\mathbb{R},\mathbb{T})$ scenario as
\begin{align}\nonumber
\text{Null}~\mathbb{EC}s:\quad&\hat{\mu}+\hat{P_r}\geq0,\quad\hat{\mu}+\hat{P}_\bot\geq0,\\\nonumber
\text{Weak}~\mathbb{EC}s:\quad&\hat{\mu}\geq0,\quad\hat{\mu}+\hat{P_r}\geq0,\quad\hat{\mu}+\hat{P}_\bot\geq0,\\\nonumber
\text{Strong}~\mathbb{EC}:\quad&\hat{\mu}+\hat{P_r}+2\hat{P}_\bot\geq0,\\\label{59}
\text{Dominant}~\mathbb{EC}s:\quad&\hat{\mu}-\hat{P}_\bot\geq0,\quad\hat{\mu}-\hat{P_r}\geq0.
\end{align}

We also discuss stability of different resulting models by making
use of multiple approaches. The causality condition requires that
the speed of sound propagation in a stable object must be less than
the speed of light, i.e., $v_{s}^2\in(0,1)$ \cite{24a}. Further,
both the components
$\big(v^2_{sr}=\frac{d\hat{P}_{r}}{d\hat{\mu}},v^2_{s\bot}=\frac{d\hat{P}_{\bot}}{d\hat{\mu}},$
refer to radial and tangential sound speed, respectively\big) in the
case of anisotropic configuration must lie within the same interval.
Herrera cracking approach reveals that the system is claimed to be
stable only if $|v^{2}_{s\bot}-v^{2}_{sr}|$ meets the above range
\cite{24}. The adiabatic index provides stable behavior for its
value greater than $\frac{4}{3}$ \cite{25}. It is defined as
\begin{equation}\label{g62}
\hat{\Gamma}=\frac{\hat{\mu}+\hat{P}_{r}}{\hat{P}_{r}}
\bigg(\frac{d\hat{P}_{r}}{d\hat{\mu}}\bigg).
\end{equation}

We investigate the developed solution in the following to check how
it behaves for different values of the parameter $\omega$.

\subsection{Radiation-Dominated Era}

In this phase, the relativistic particles such as neutrinos and
photons formed matter that dominated our cosmos. This era can be
discussed by inserting $\omega=\frac{1}{3}$ in
$\mathbb{E}o\mathbb{S}$ \eqref{16a}, which implies that pressure is
equivalent to one by three times the entire energy density. Also,
the mass of the particles within the universe is observed to be less
than their momentum. Figure \textbf{1} displays the graphs of
physical variables and anisotropy for both values of the decoupling
parameter $\delta$. The energy density (upper left plot) depicts
rapidly decreasing behavior with the increment in time, which
suggests cosmic expansion. We also find a less dense profile of this
model with the rise in $\delta$, whereas the behavior of pressure
component is shown to be opposite in this scenario. The lower right
plot indicates that the radial pressure is slightly greater than the
tangential one (negative anisotropy) only at initial time, and then
anisotropy becomes positive. All the $\mathbb{EC}s$ are plotted in
Figure \textbf{2} which shows a viable cosmic model. Figure
\textbf{3} expresses that all the stability criteria are fulfilled,
thus this extended solution portrays stable geometry for
$\omega=\frac{1}{3}$.
\begin{figure}\center
\epsfig{file=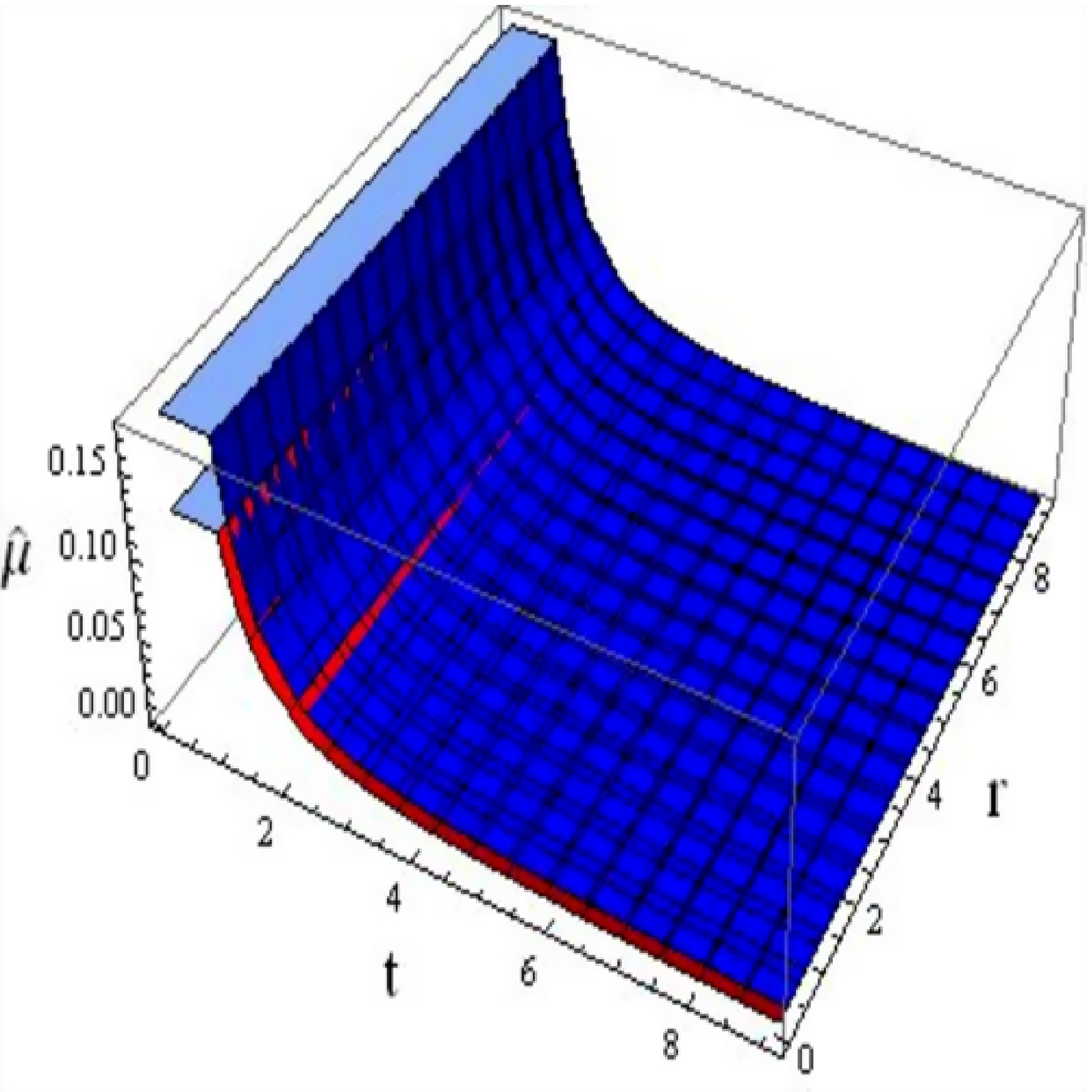,width=0.4\linewidth}\epsfig{file=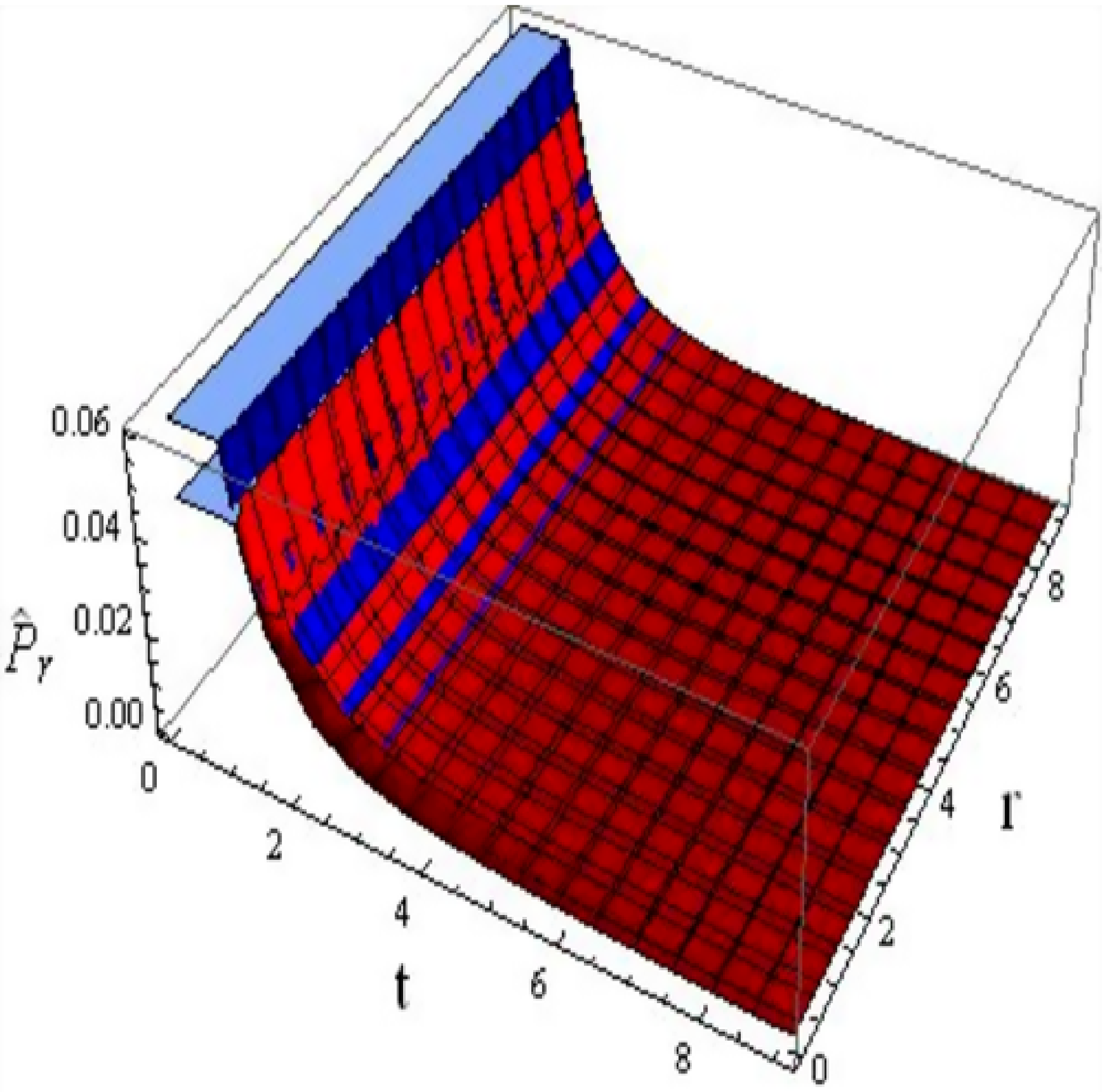,width=0.4\linewidth}
\epsfig{file=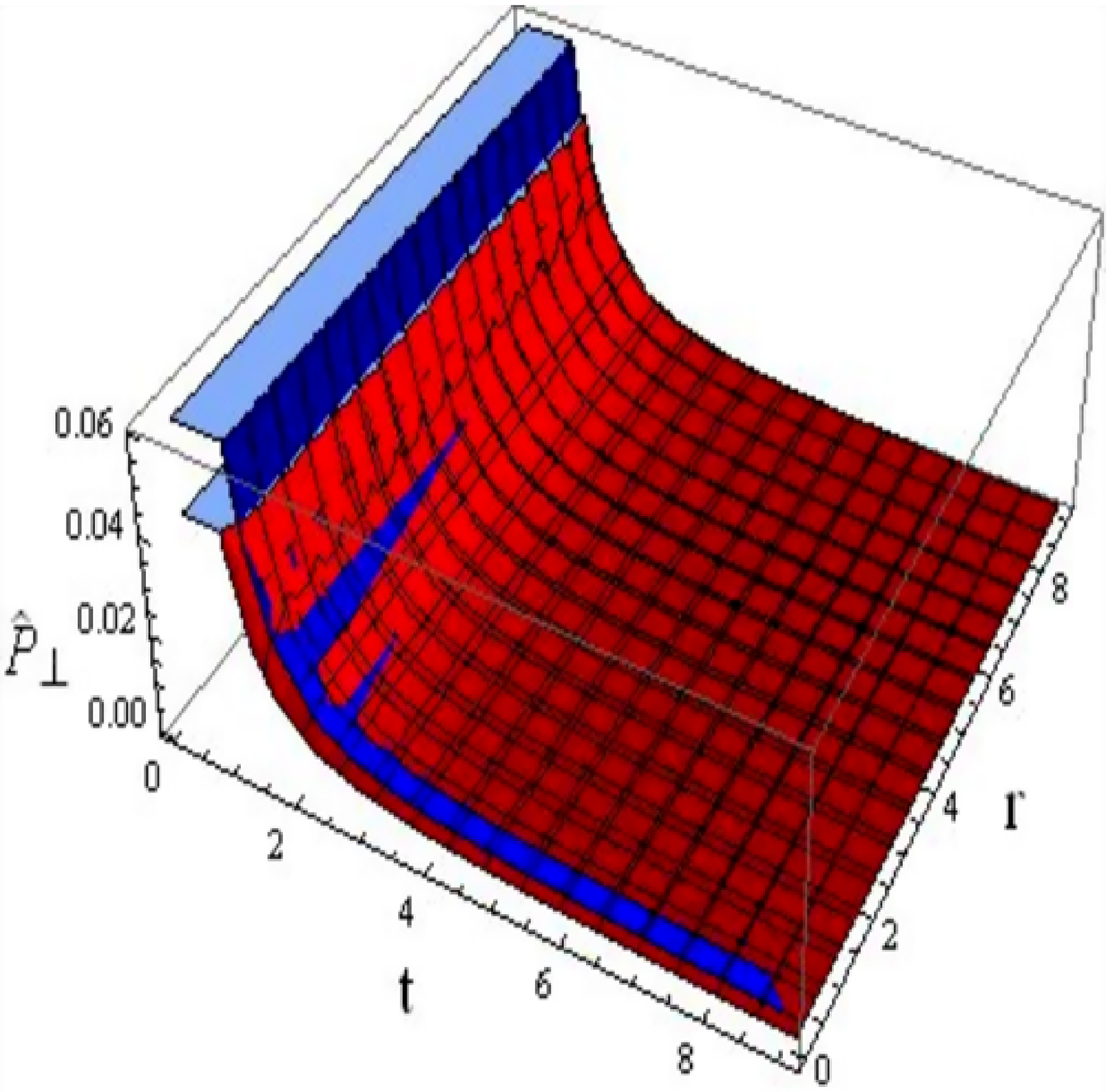,width=0.4\linewidth}\epsfig{file=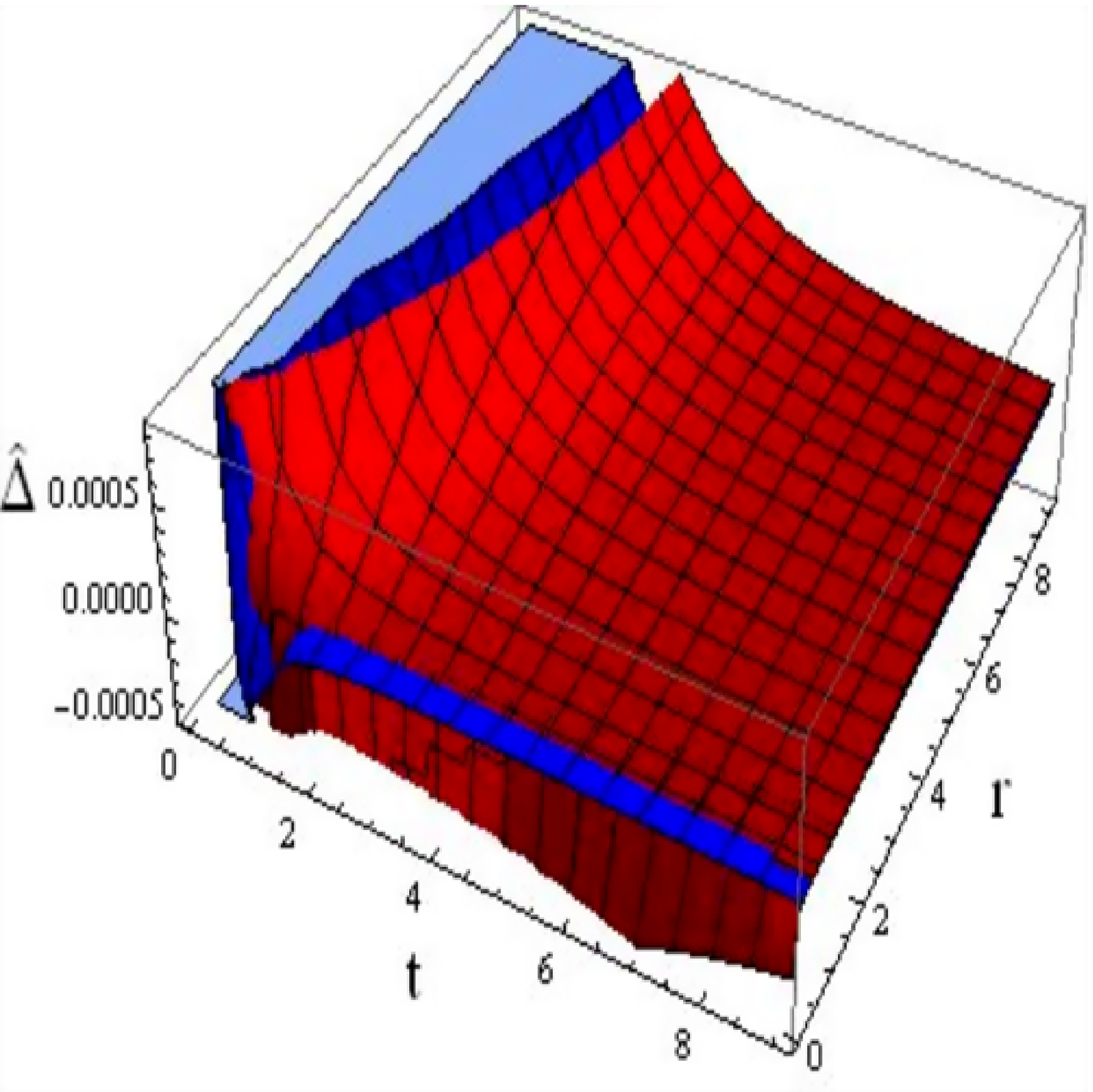,width=0.4\linewidth}
\caption{Plots of matter variables and anisotropy with
$\omega=\frac{1}{3}$ for $\delta=0.1$ (blue) and $0.9$ (red).}
\end{figure}
\begin{figure}\center
\epsfig{file=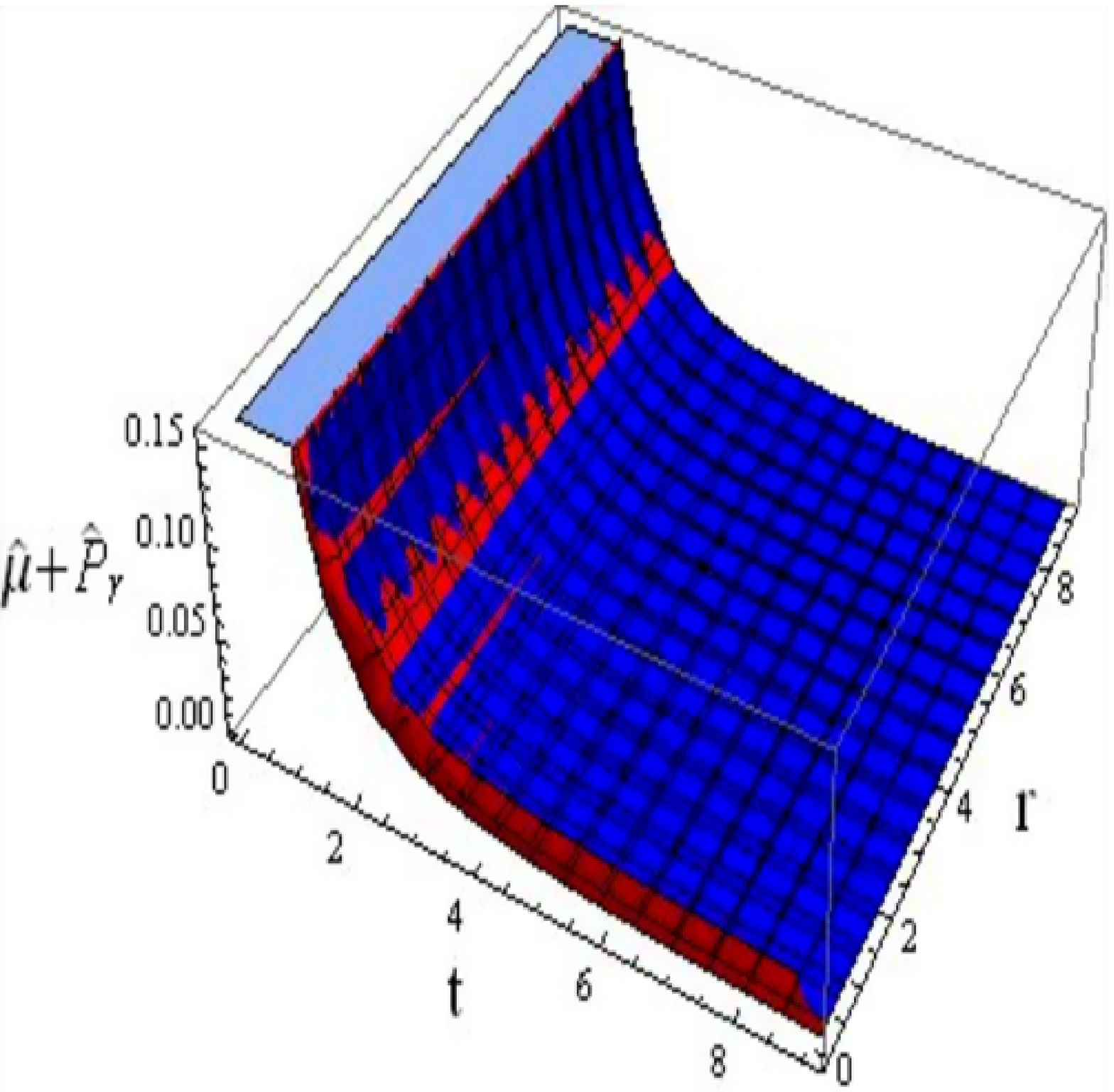,width=0.4\linewidth}\epsfig{file=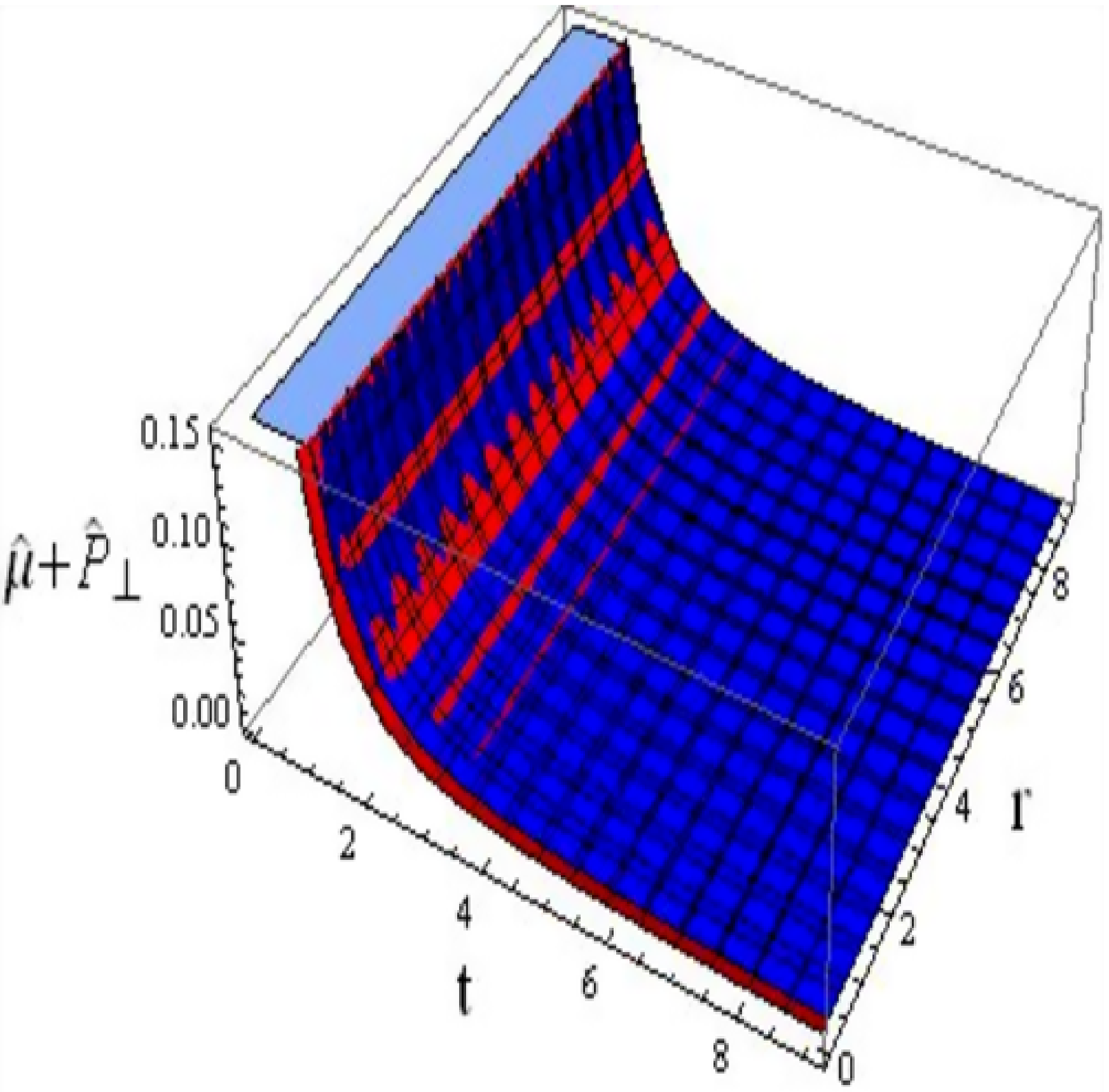,width=0.4\linewidth}
\epsfig{file=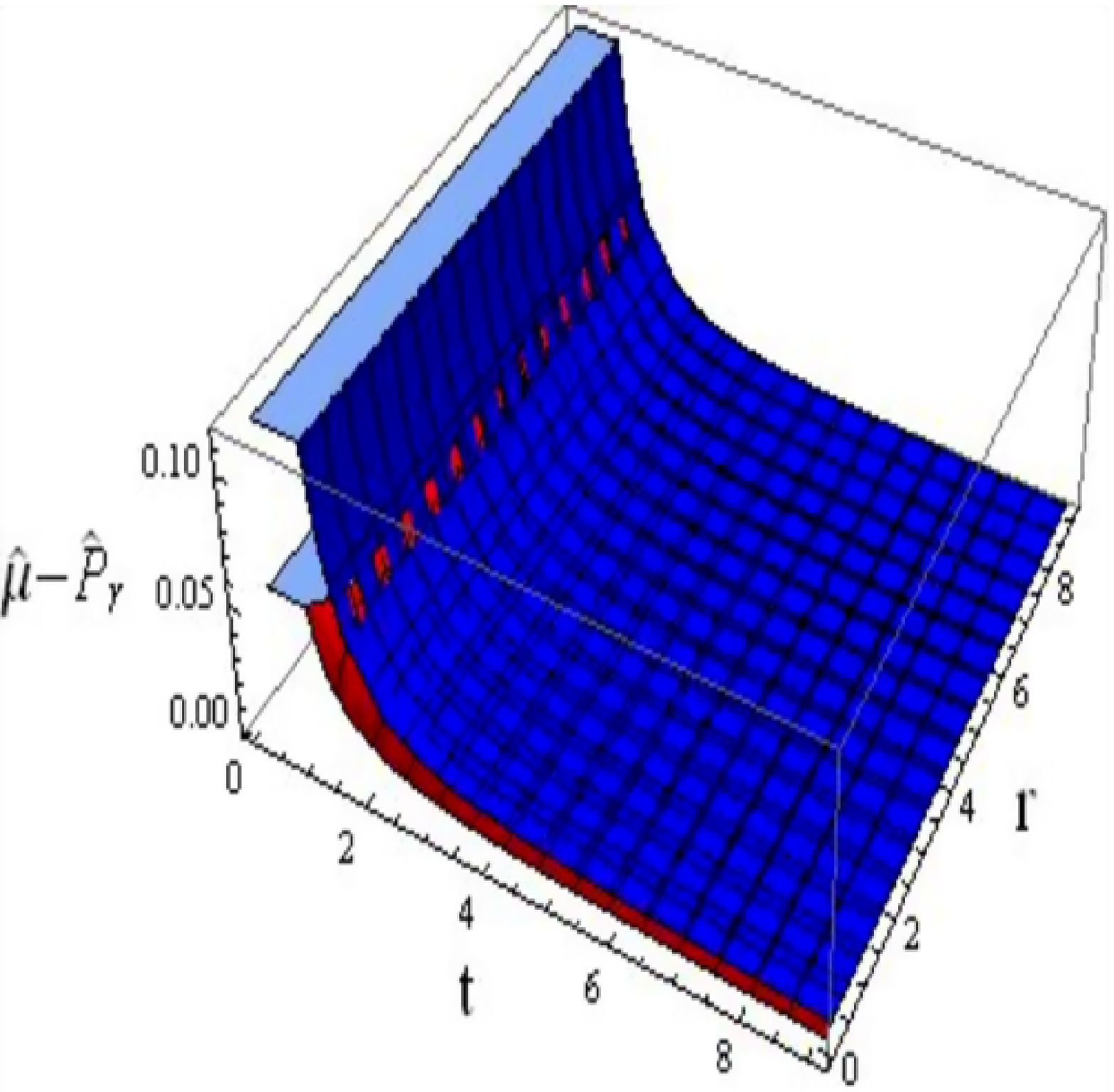,width=0.4\linewidth}\epsfig{file=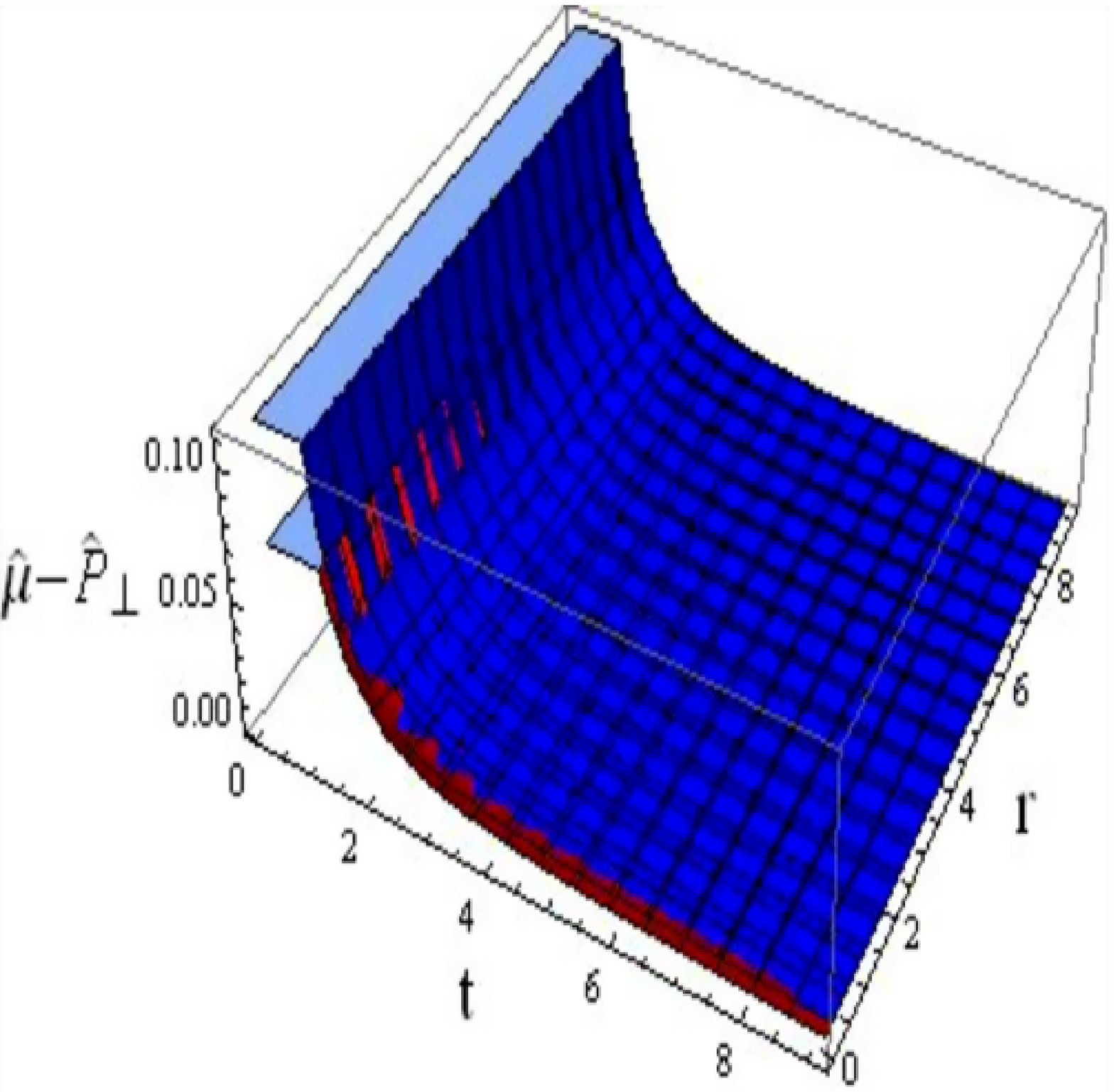,width=0.4\linewidth}
\epsfig{file=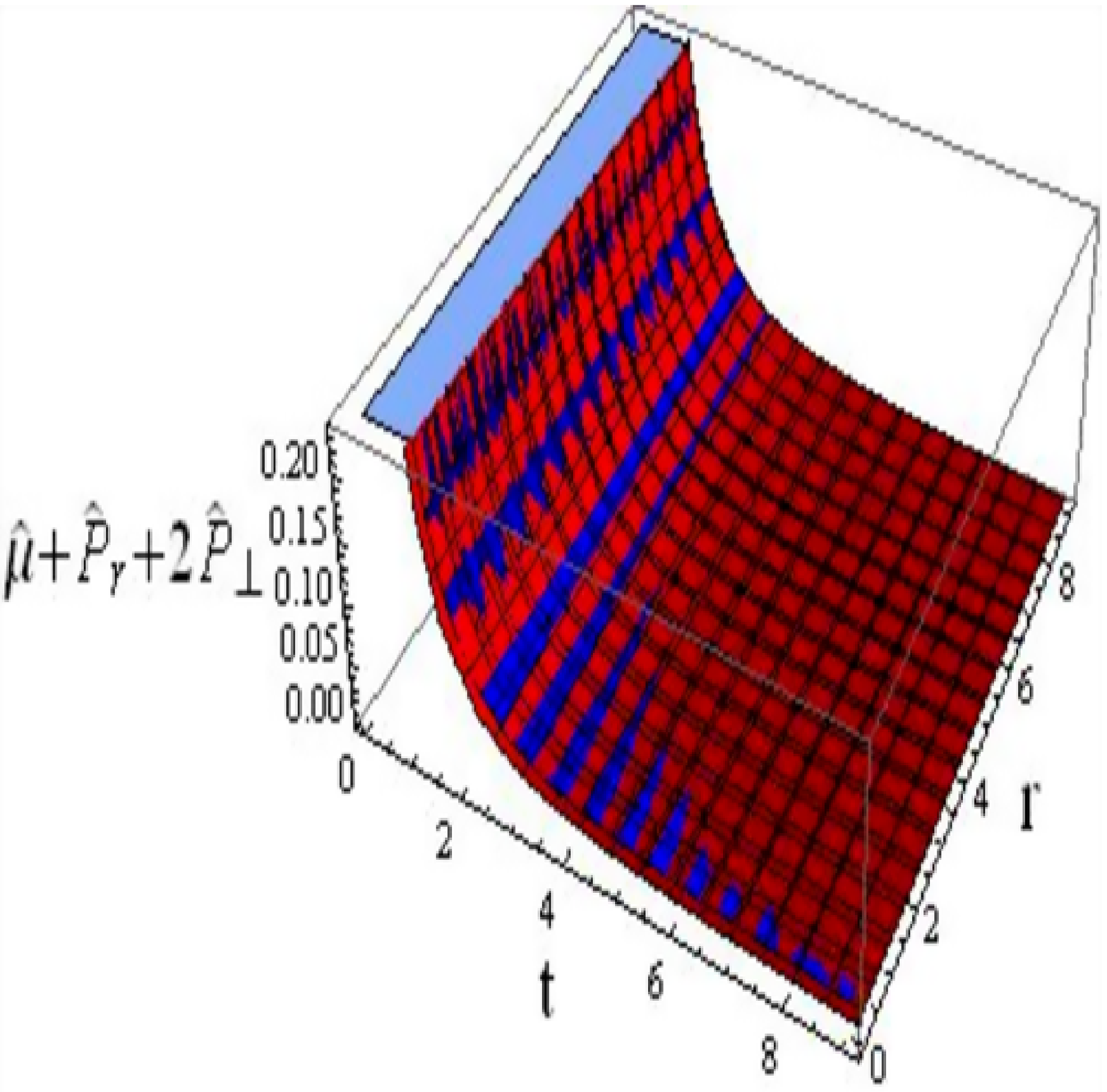,width=0.4\linewidth} \caption{Plots of
$\mathbb{EC}s$ with $\omega=\frac{1}{3}$ for $\delta=0.1$ (blue) and
$0.9$ (red).}
\end{figure}
\begin{figure}\center
\epsfig{file=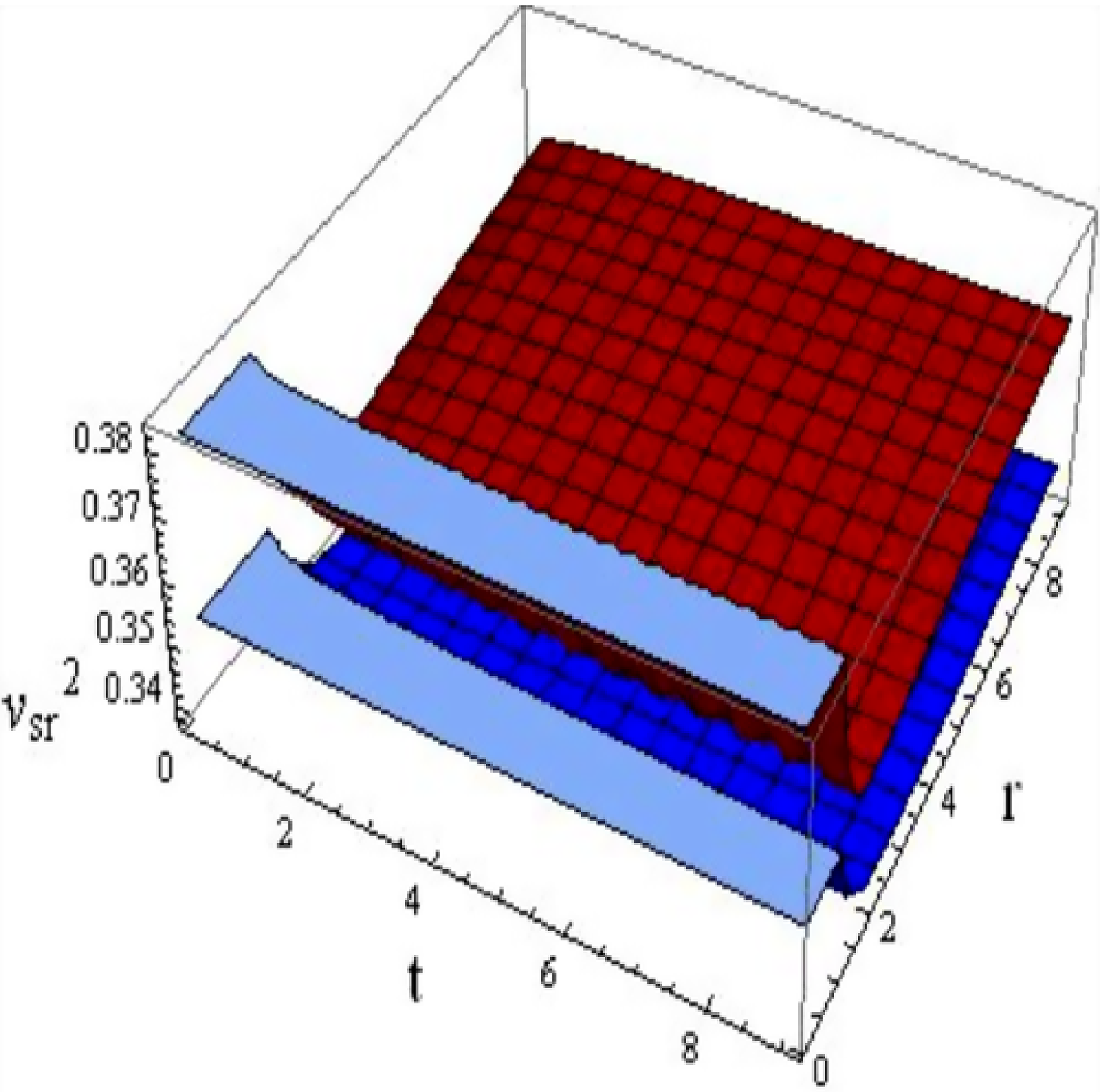,width=0.4\linewidth}\epsfig{file=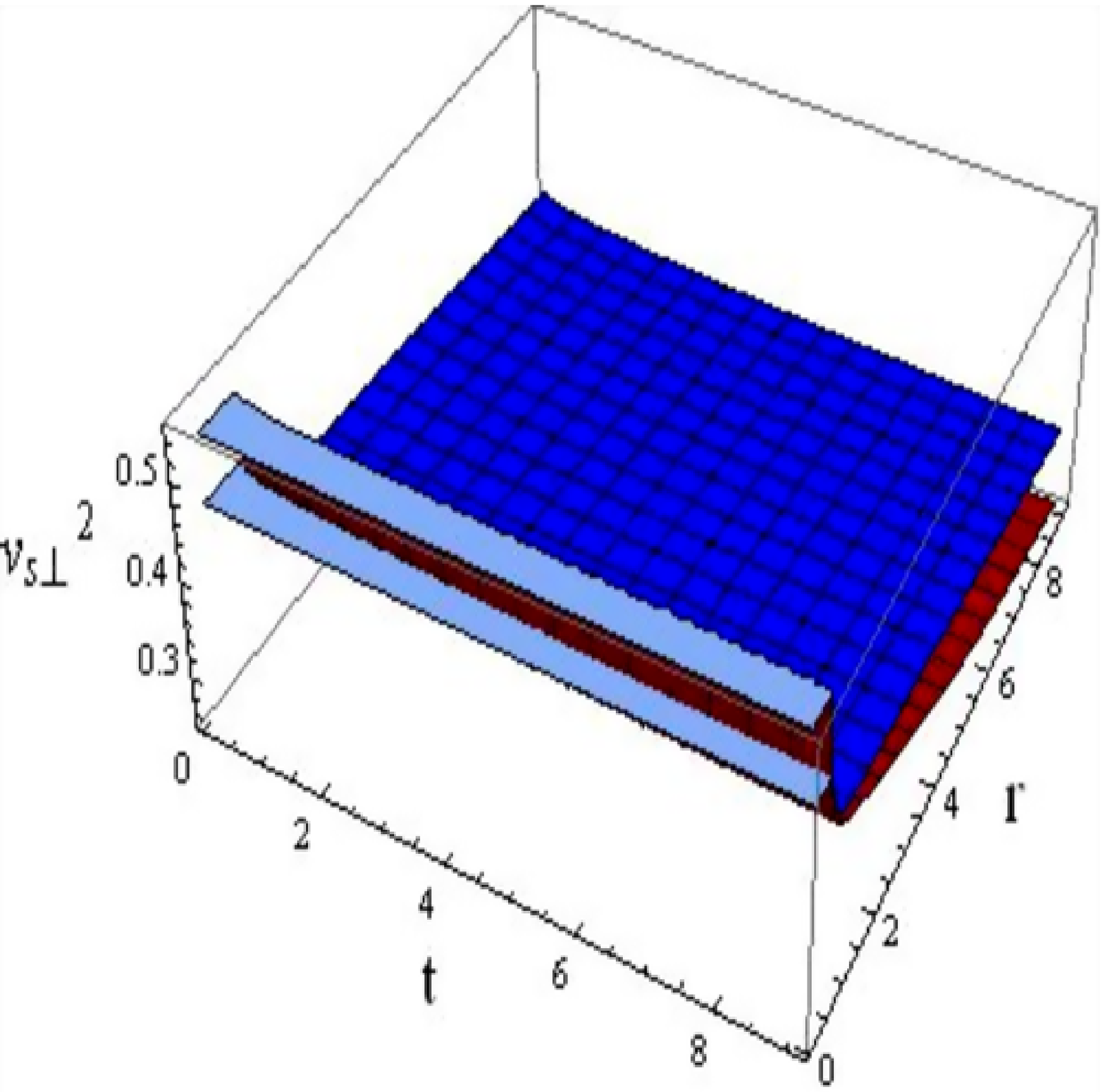,width=0.4\linewidth}
\epsfig{file=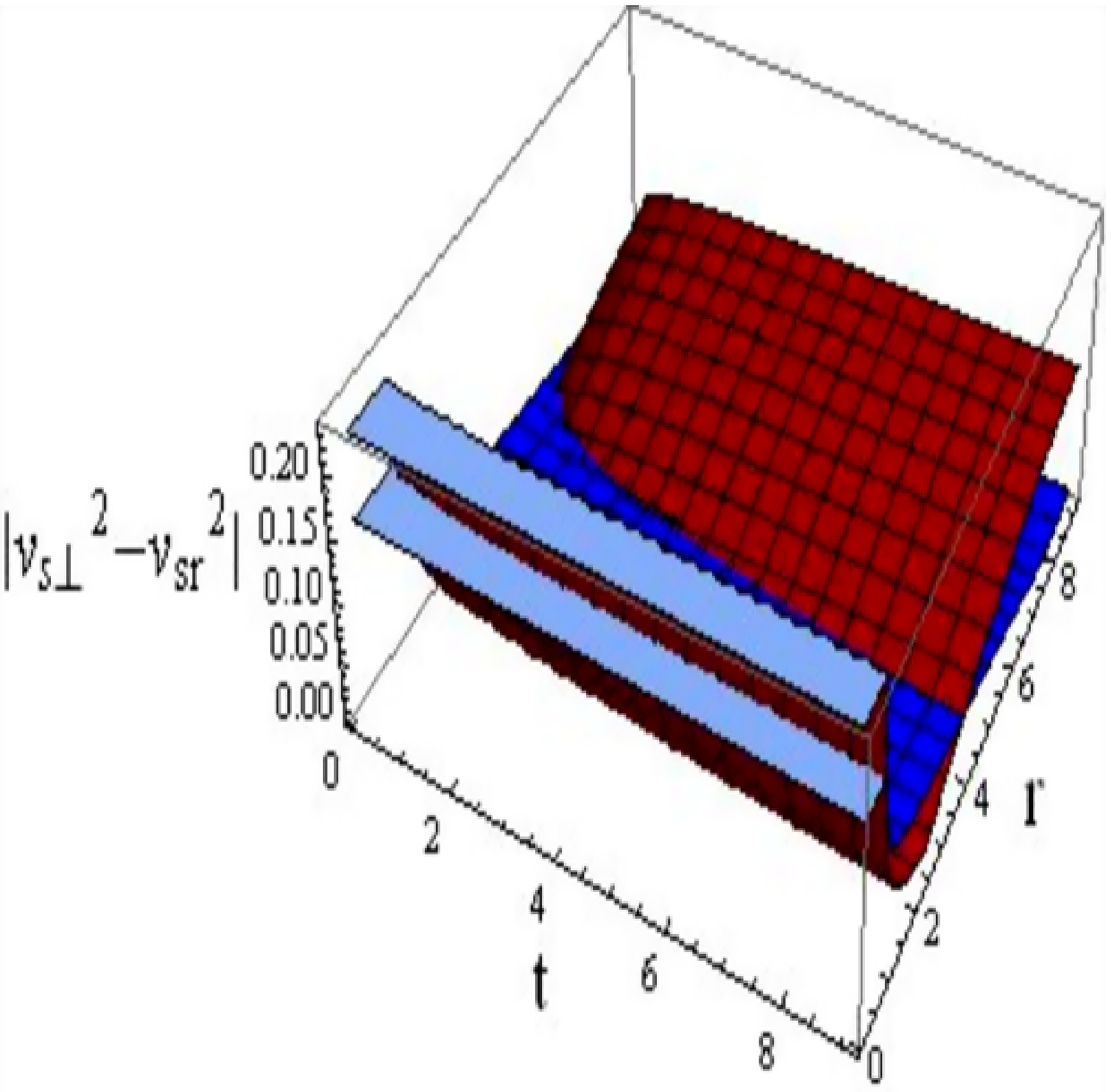,width=0.4\linewidth}\epsfig{file=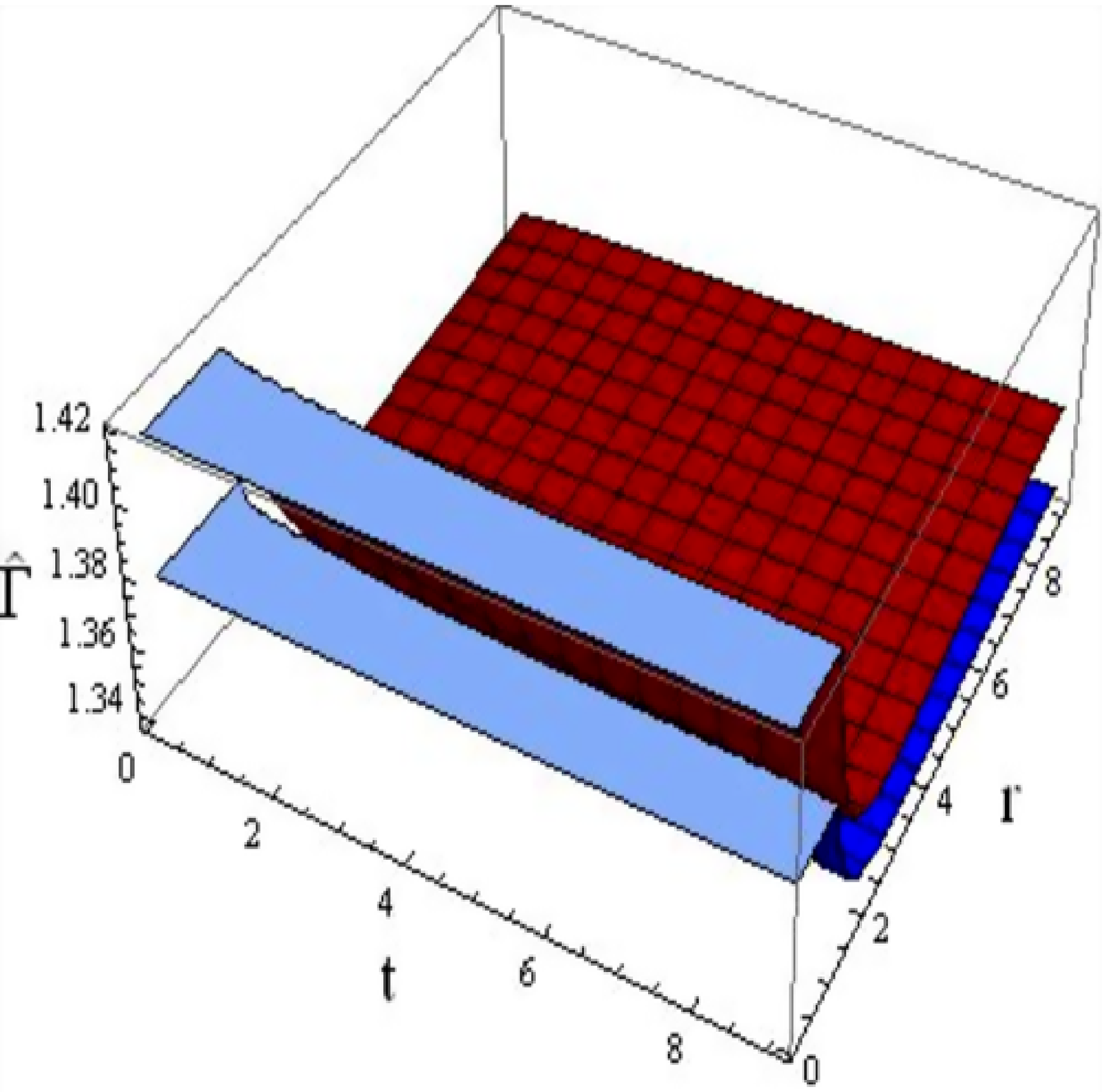,width=0.4\linewidth}
\caption{Plots of $v^{2}_{sr}$, $v^{2}_{s\bot}$,
$|v^{2}_{s\bot}-v^{2}_{sr}|$ and adiabatic index with
$\omega=\frac{1}{3}$ for $\delta=0.1$ (blue) and $0.9$ (red).}
\end{figure}

\subsection{Matter-Dominated Era}

A large portion of our cosmos encompasses non-relativistic particles
(baryons) during this phase. These are the particles having mass
energy greater than their kinetic energy. We can define dust as a
substance that is made up of such elementary particles. Figure
\textbf{4} interprets physical determinants and the pressure
anisotropy. The energy density displays decreasing trend with time
(showing cosmic expansion) and observed to be in an inverse relation
with the parameter $\delta$ (upper left). As $\omega=0$ leads to the
matter-dominated era, resulting in the vanishing isotropic pressure
that can be observed from $\mathbb{E}o\mathbb{S}$ \eqref{16a}, thus
we term it the dust era. However, the radial/tangential pressures do
not vanish individually in this case (Figure \textbf{4}) but the
corresponding isotropic pressure \big(i.e.,
$P=\frac{\hat{P}_r+2\hat{P}_\bot}{3}$\big) would vanish. Figure
\textbf{5} verifies fulfillment of all the $\mathbb{EC}s$ that
results in a viable model. We also obtain stability of the solution
\eqref{42}-\eqref{42c} corresponding to $\omega=0$ (Figure
\textbf{6}).
\begin{figure}\center
\epsfig{file=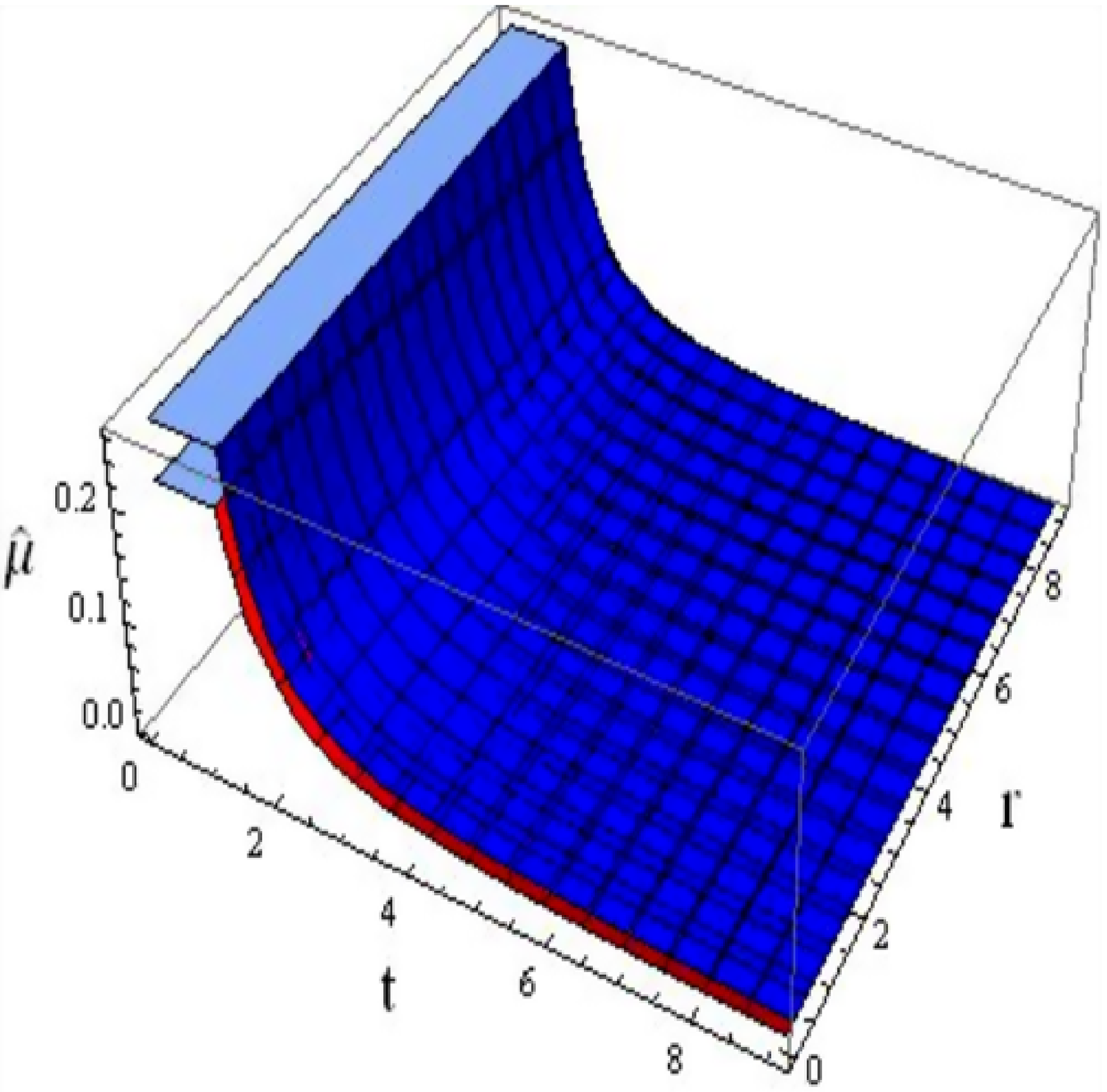,width=0.4\linewidth}\epsfig{file=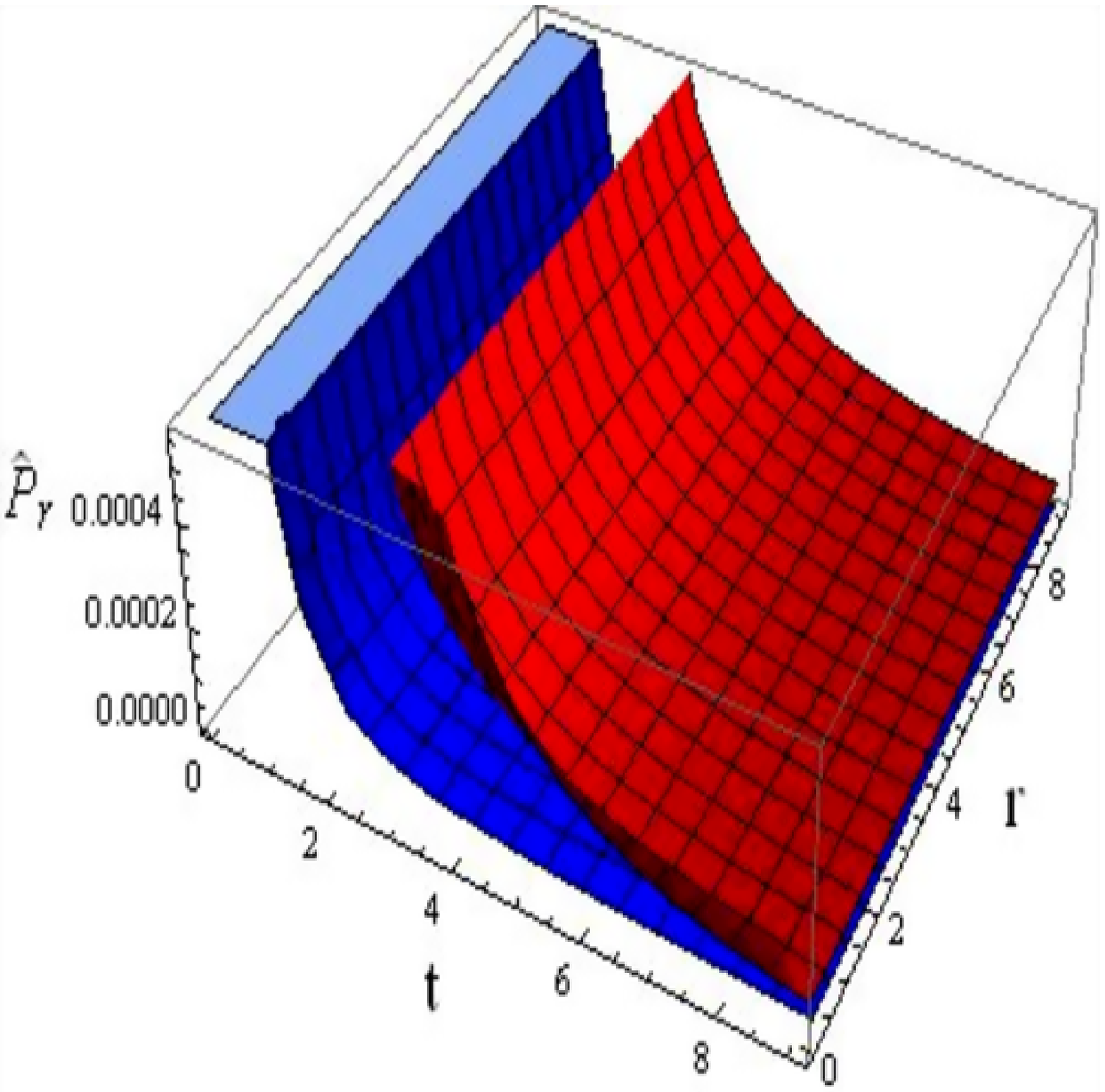,width=0.4\linewidth}
\epsfig{file=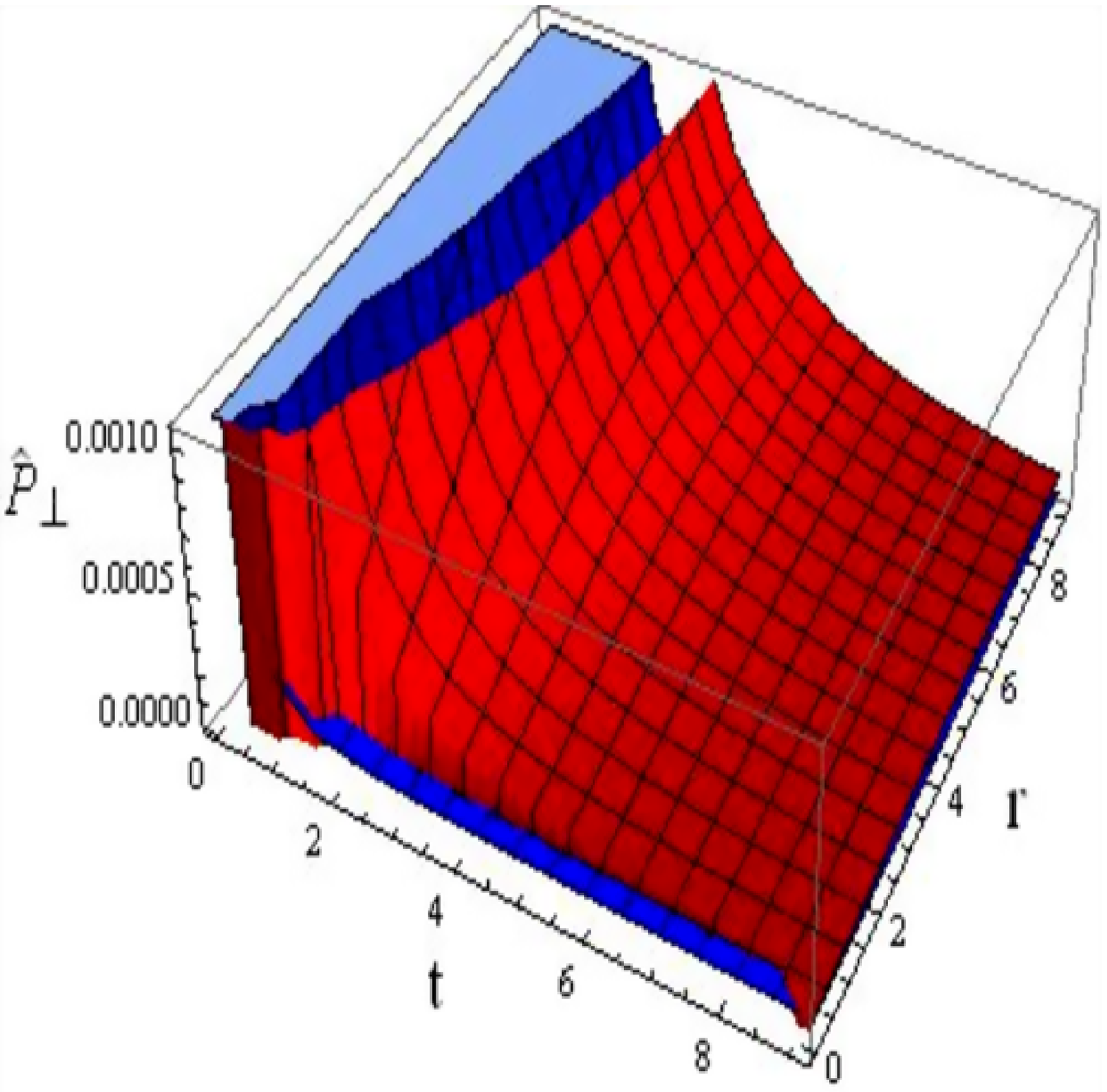,width=0.4\linewidth}\epsfig{file=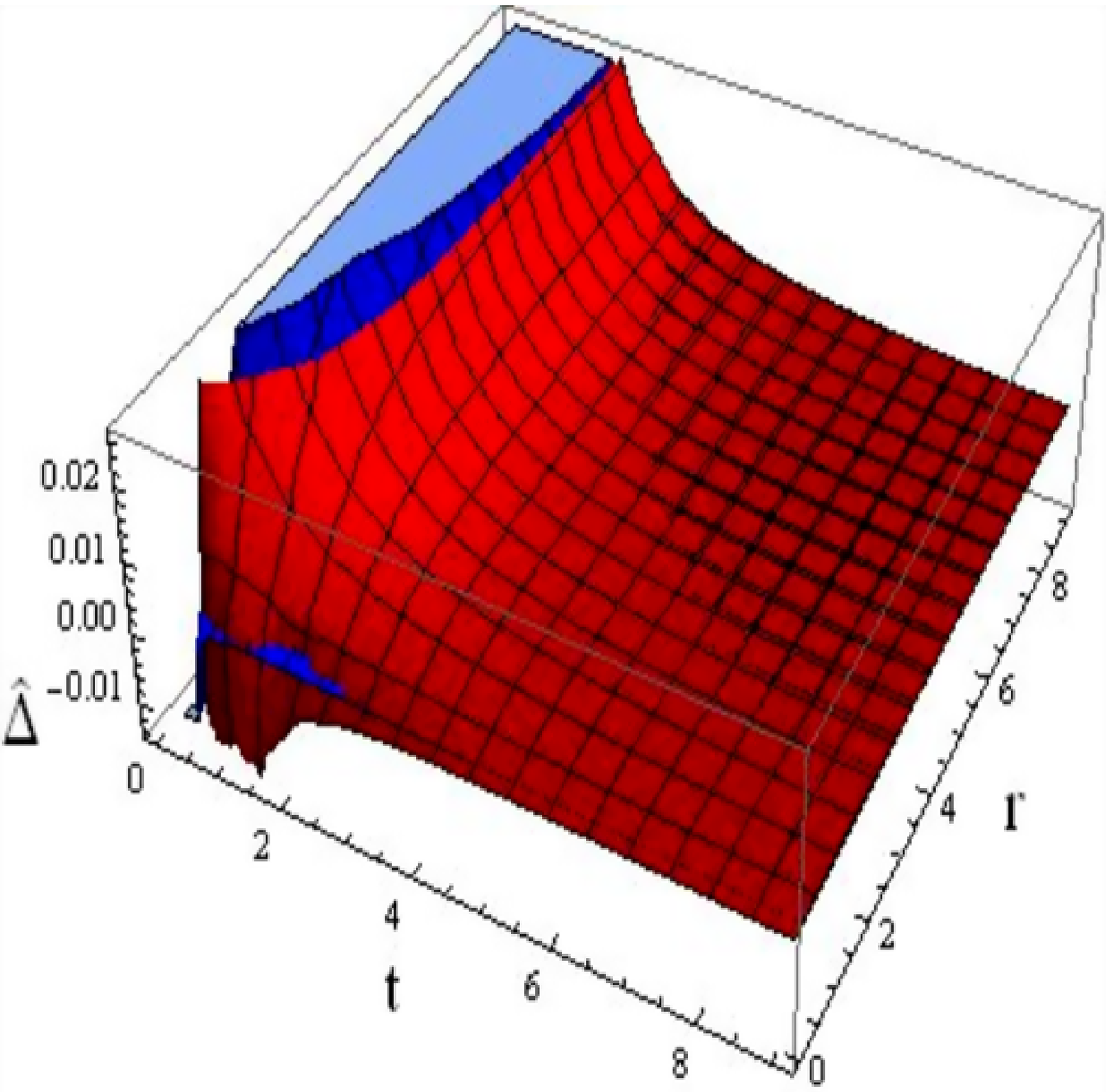,width=0.4\linewidth}
\caption{Plots of matter variables and anisotropy with $\omega=0$
for $\delta=0.1$ (blue) and $0.9$ (red).}
\end{figure}
\begin{figure}\center
\epsfig{file=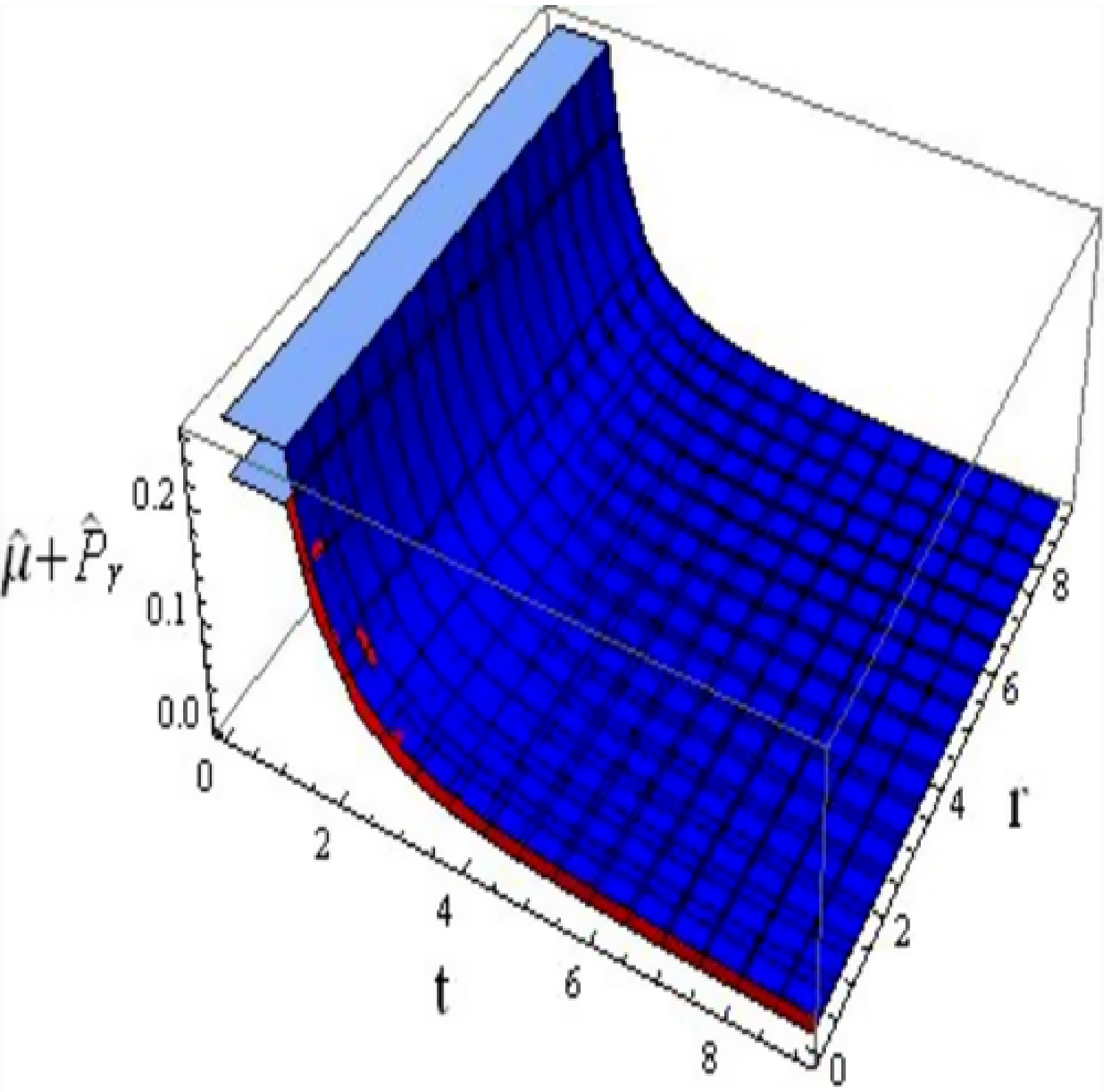,width=0.4\linewidth}\epsfig{file=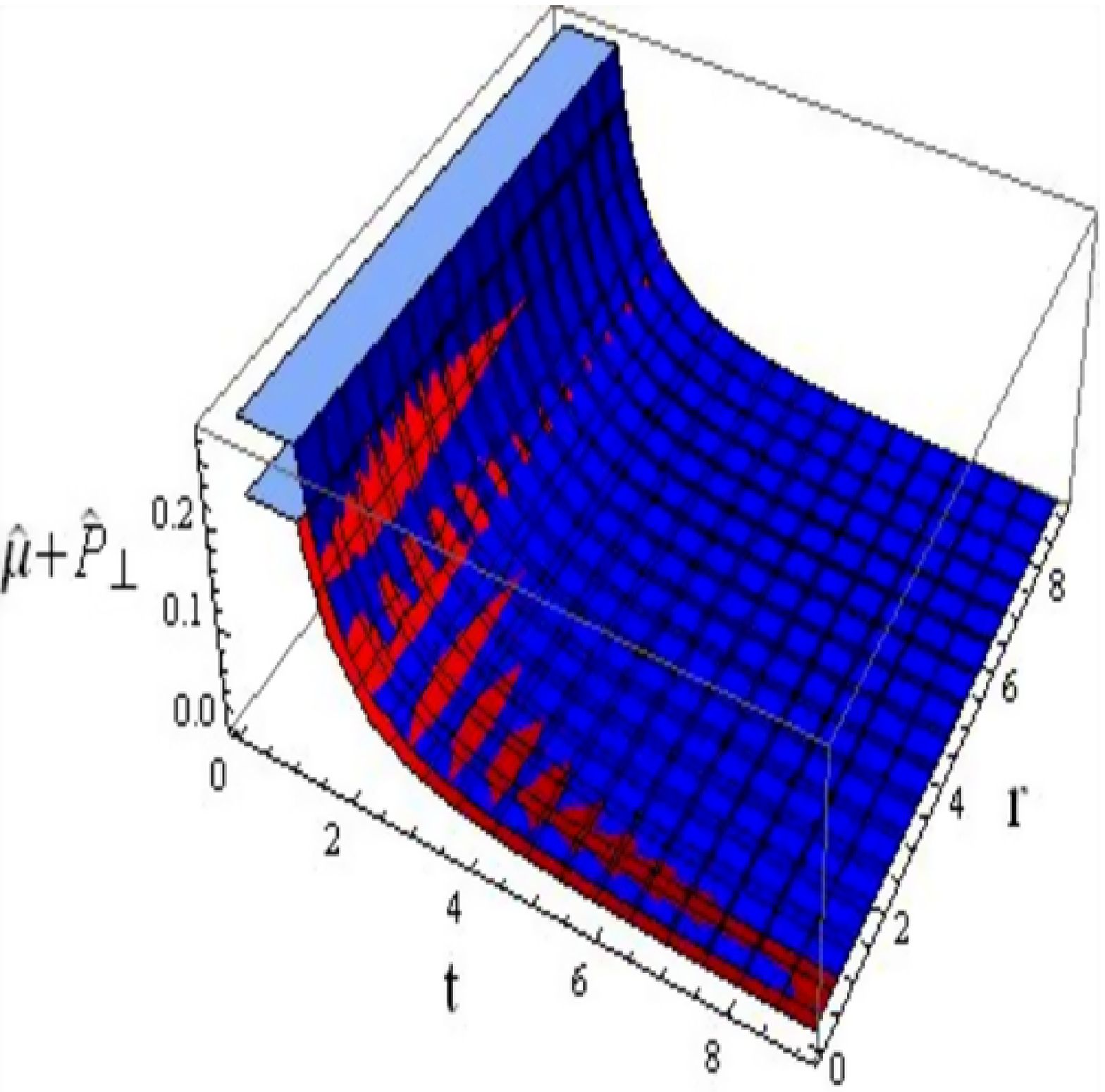,width=0.4\linewidth}
\epsfig{file=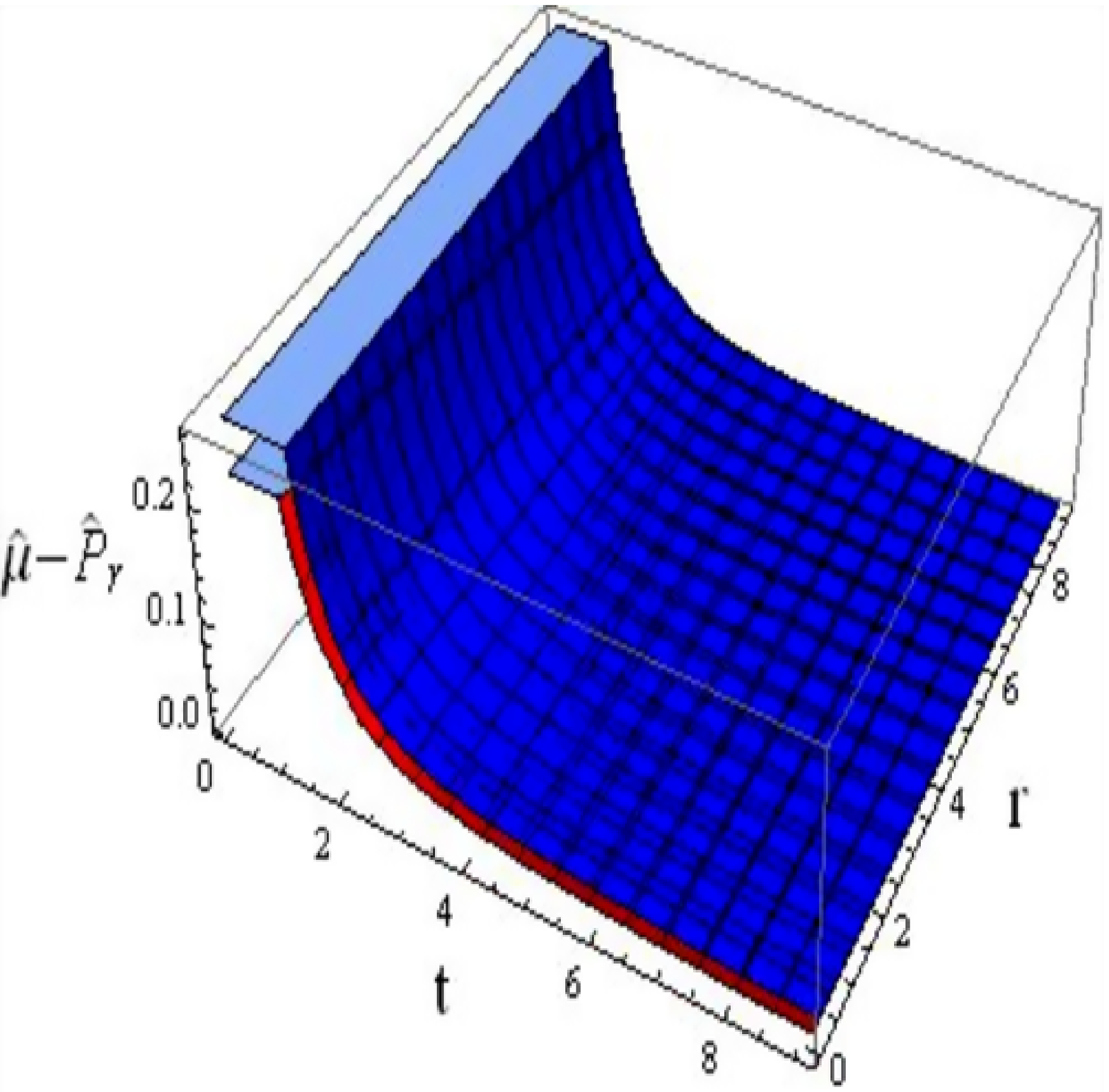,width=0.4\linewidth}\epsfig{file=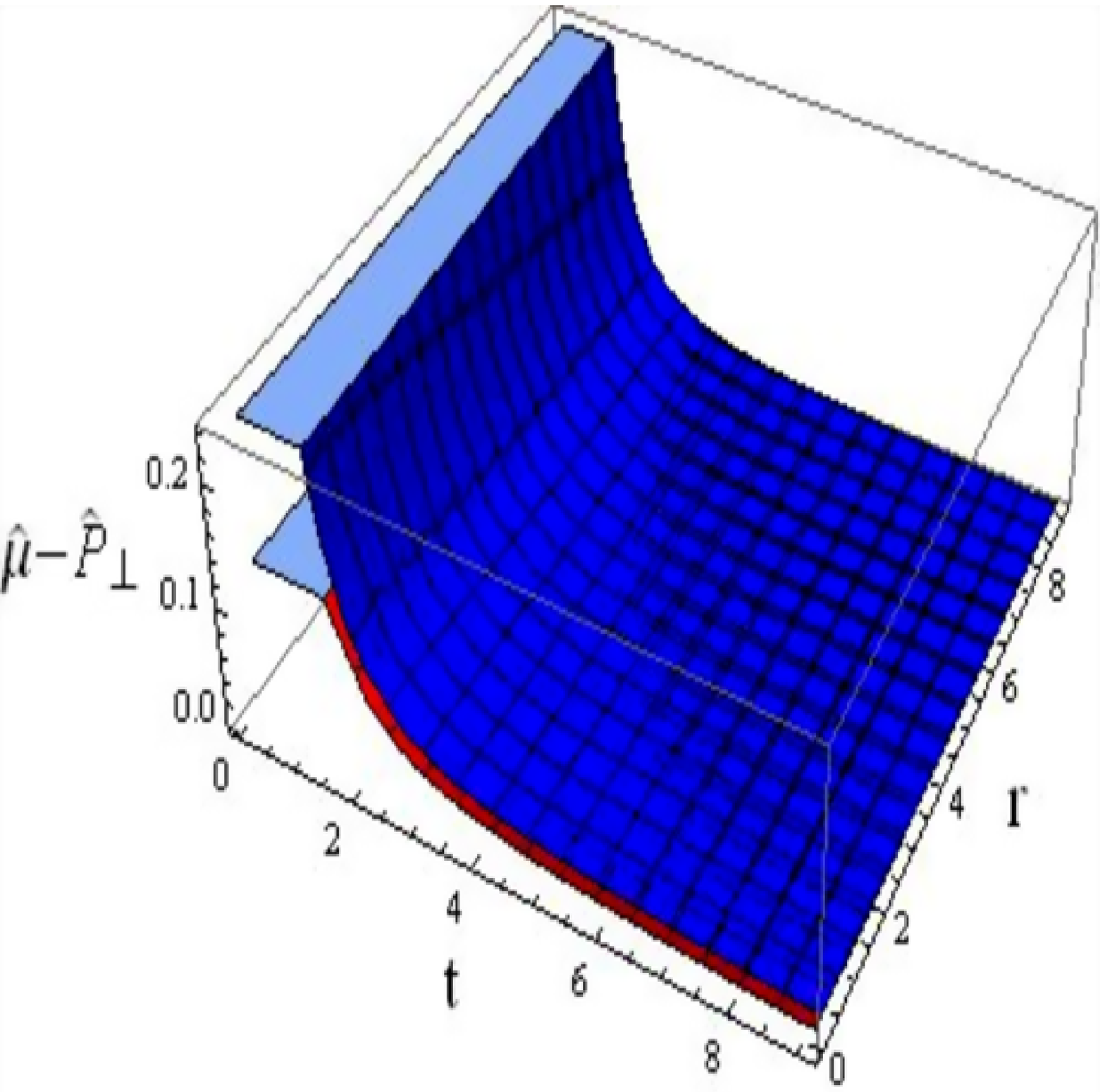,width=0.4\linewidth}
\epsfig{file=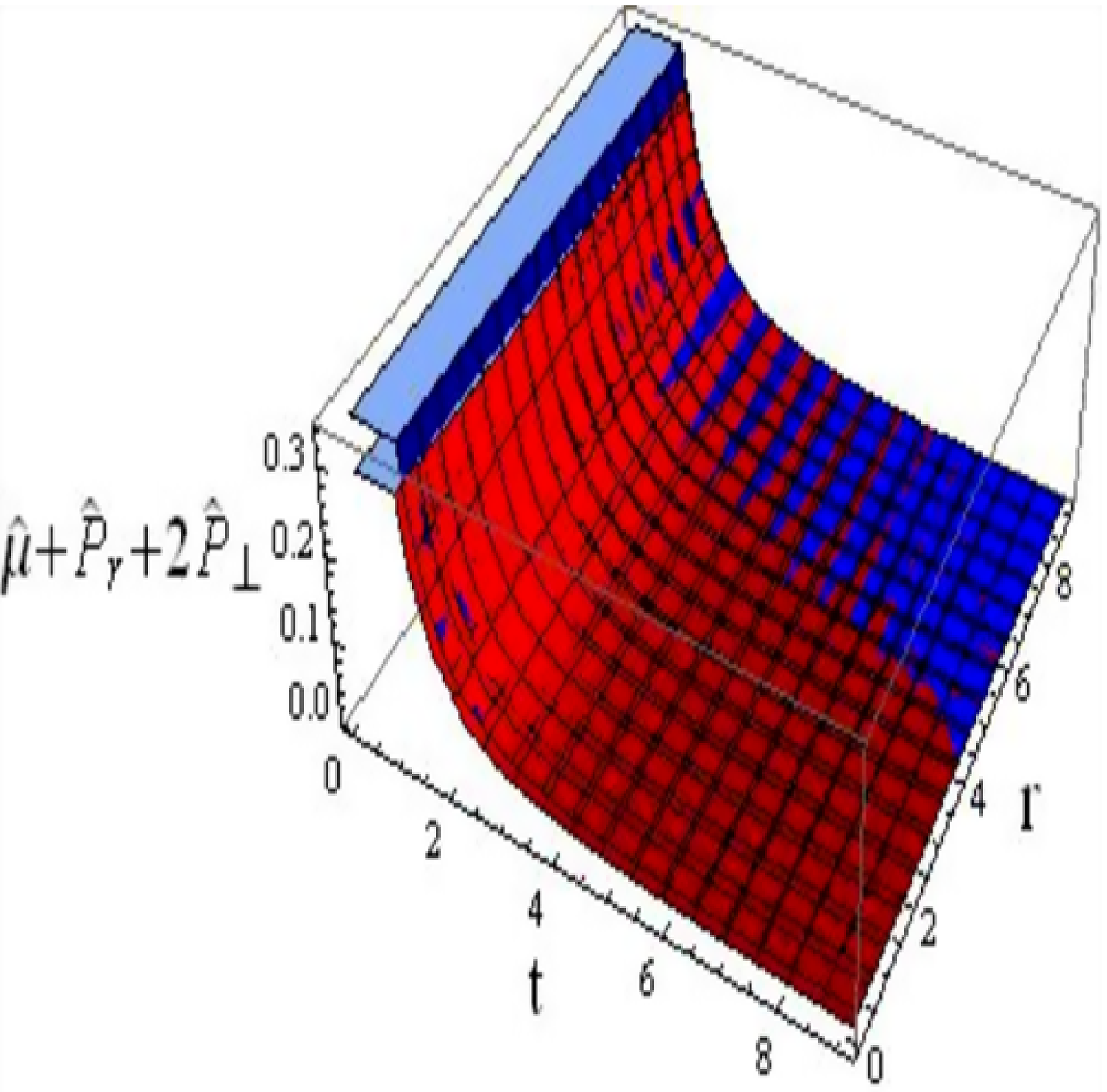,width=0.4\linewidth} \caption{Plots of
$\mathbb{EC}s$ with $\omega=0$ for $\delta=0.1$ (blue) and $0.9$
(red).}
\end{figure}
\begin{figure}\center
\epsfig{file=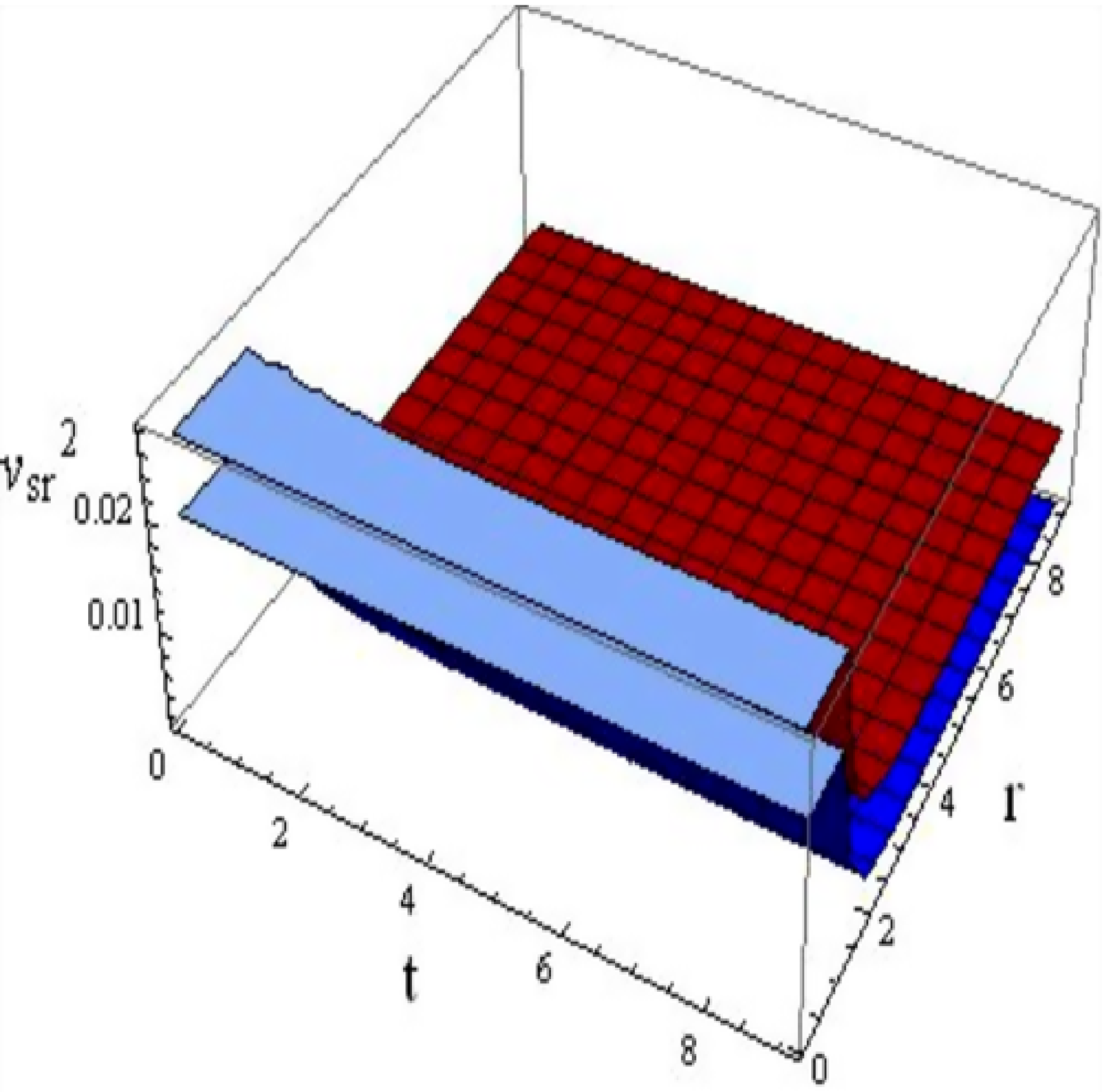,width=0.4\linewidth}\epsfig{file=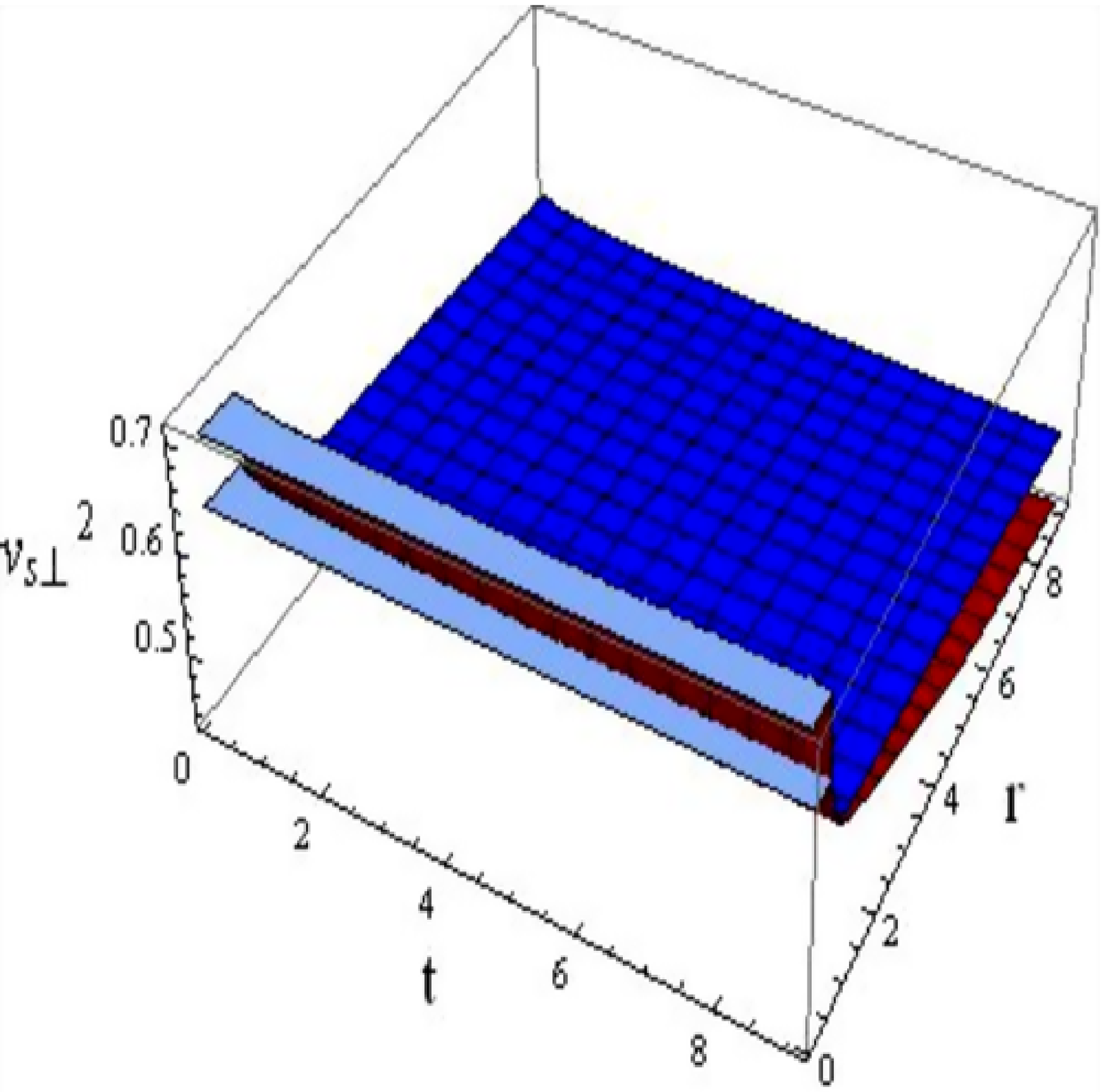,width=0.4\linewidth}
\epsfig{file=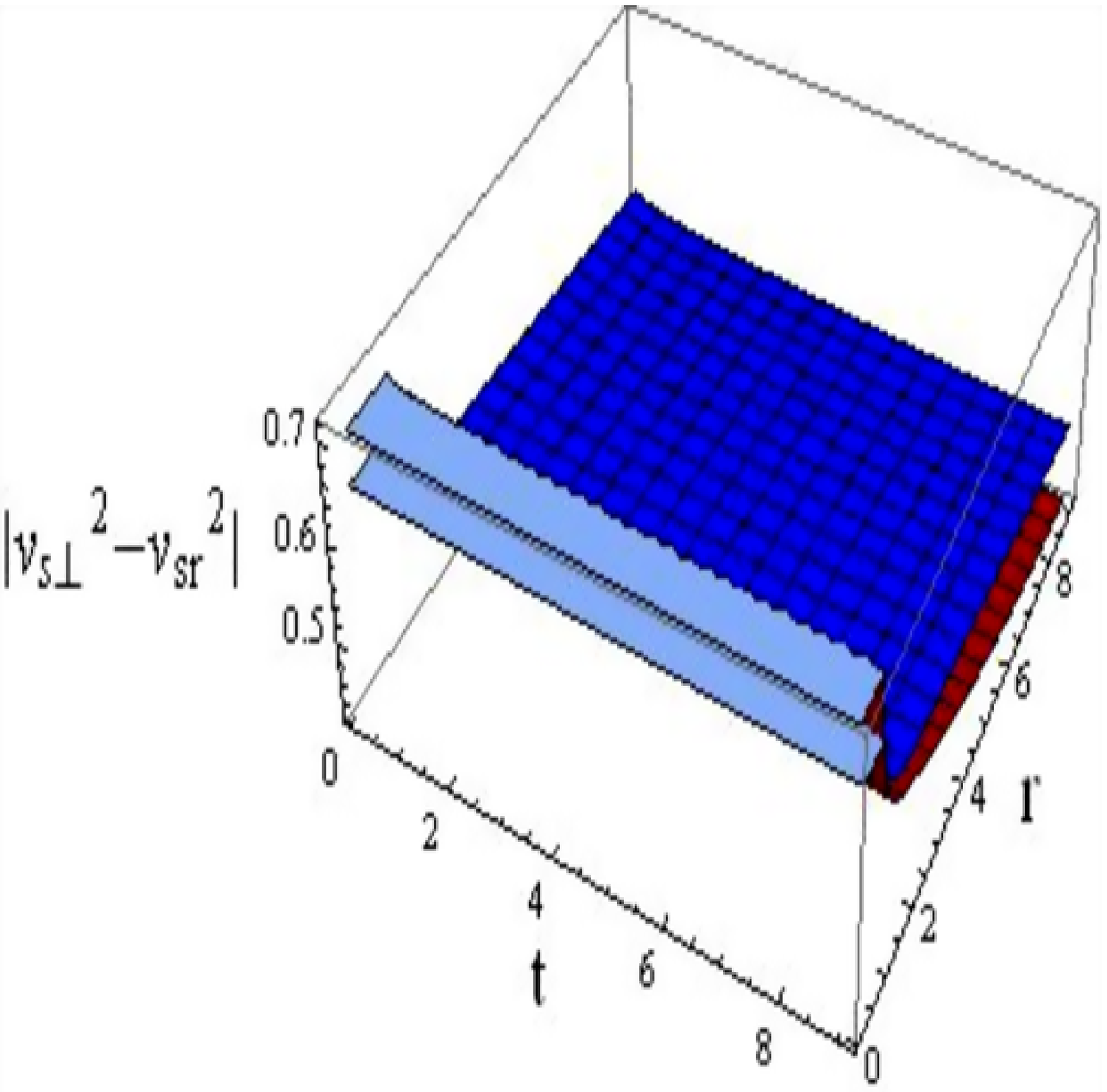,width=0.4\linewidth}\epsfig{file=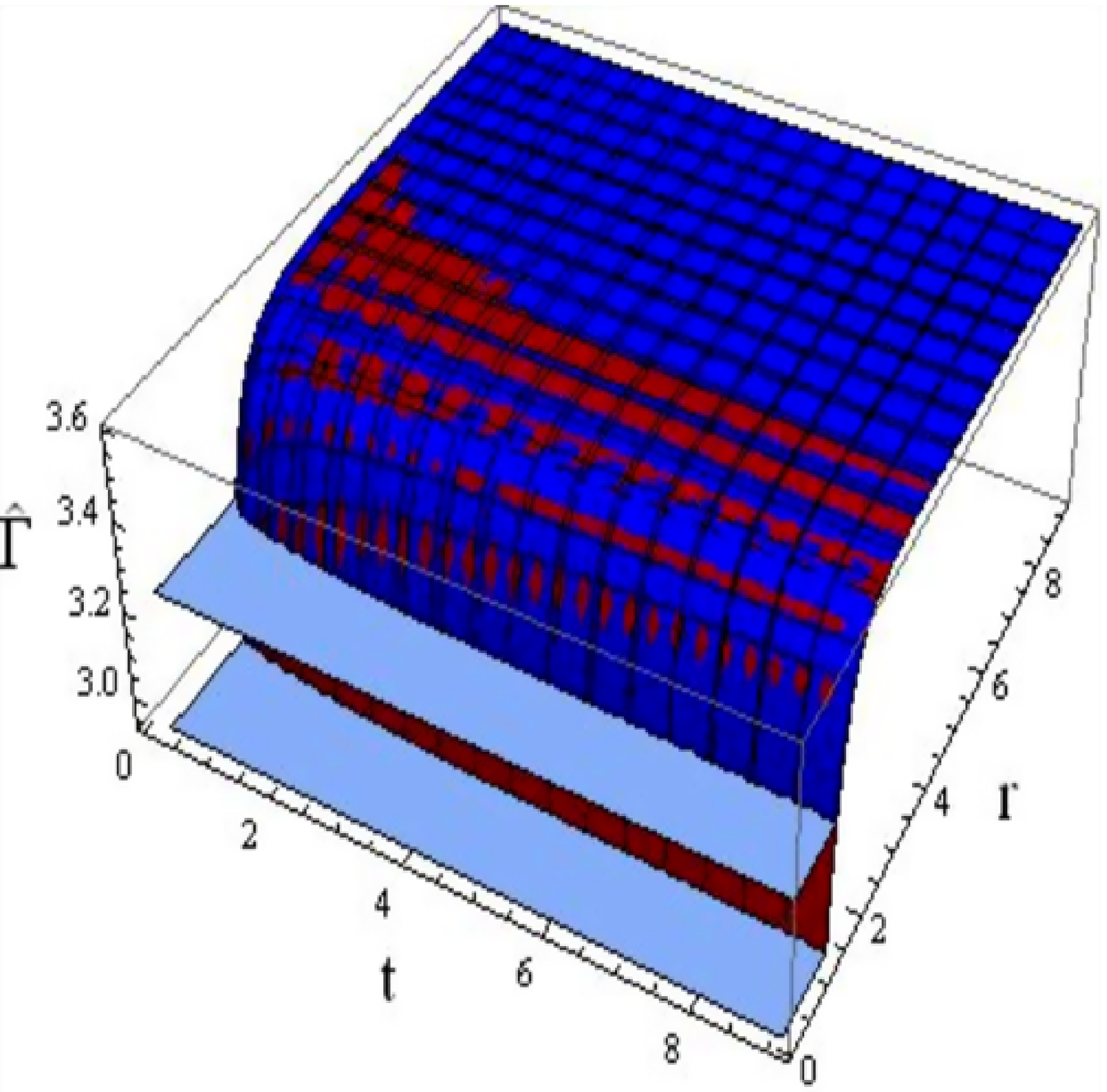,width=0.4\linewidth}
\caption{Plots of $v^{2}_{sr}$, $v^{2}_{s\bot}$,
$|v^{2}_{s\bot}-v^{2}_{sr}|$ and adiabatic index with $\omega=0$ for
$\delta=0.1$ (blue) and $0.9$ (red).}
\end{figure}

\subsection{Vacuum Energy Dominated Era}

This era is acknowledged as the last in cosmic evolutionary phases
which can be examined for $\omega=-1$. The vacuum energy (or dark
energy) dominates the matter density in this epoch, and consequently
this universe expands at an acceleration rate. Figure \textbf{7}
demonstrates the graphical nature of state variables and the
anisotropic factor. The density profile corresponding to this era is
observed to be the same with respect to $\delta$ as we have obtained
for the previous epochs. The negative trend of radial and tangential
pressures confirm the existence of a repulsive force that increases
the rate of expansion. Moreover, the increment in the parameter
$\delta$ makes the expansion rate more faster. The null, weak and
strong $\mathbb{EC}s$ are violated in this case, thus the resulting
model is not viable (Figure \textbf{8}). The plots in Figure
\textbf{9} are not consistent with their respective criteria for
both values of $\delta$, thus the extended anisotropic solution is
unstable everywhere for $\omega=-1$.
\begin{figure}\center
\epsfig{file=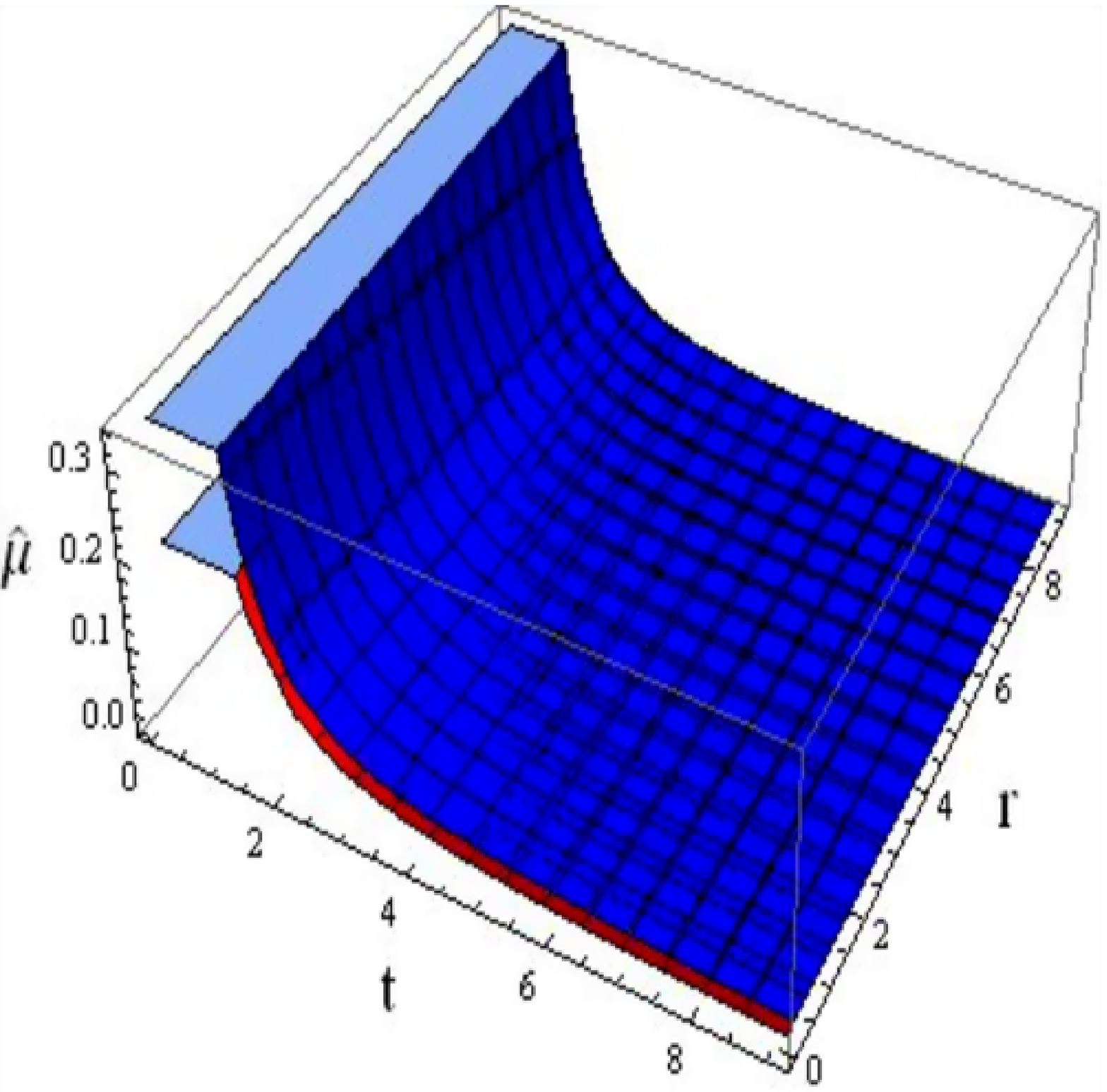,width=0.4\linewidth}\epsfig{file=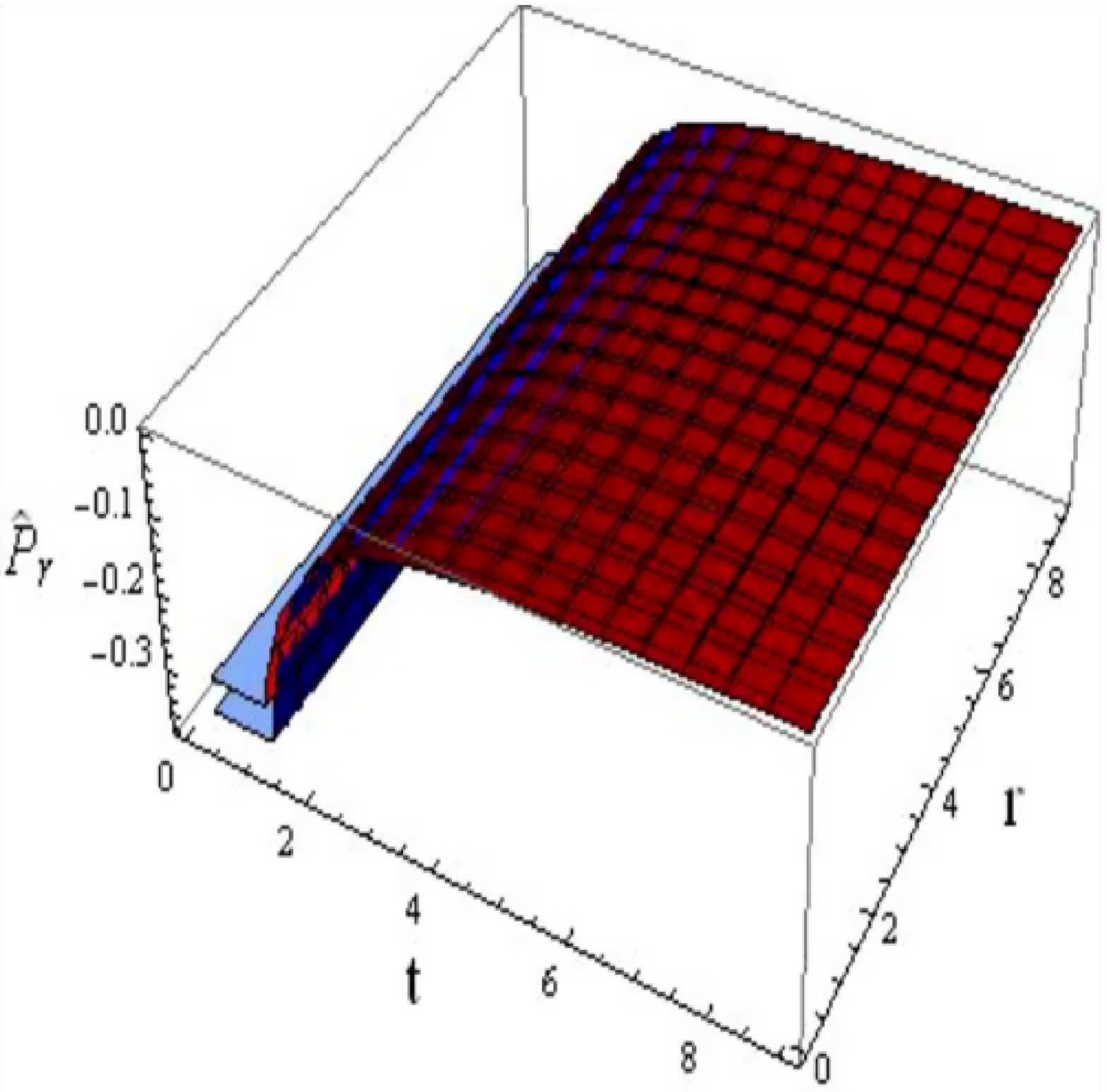,width=0.4\linewidth}
\epsfig{file=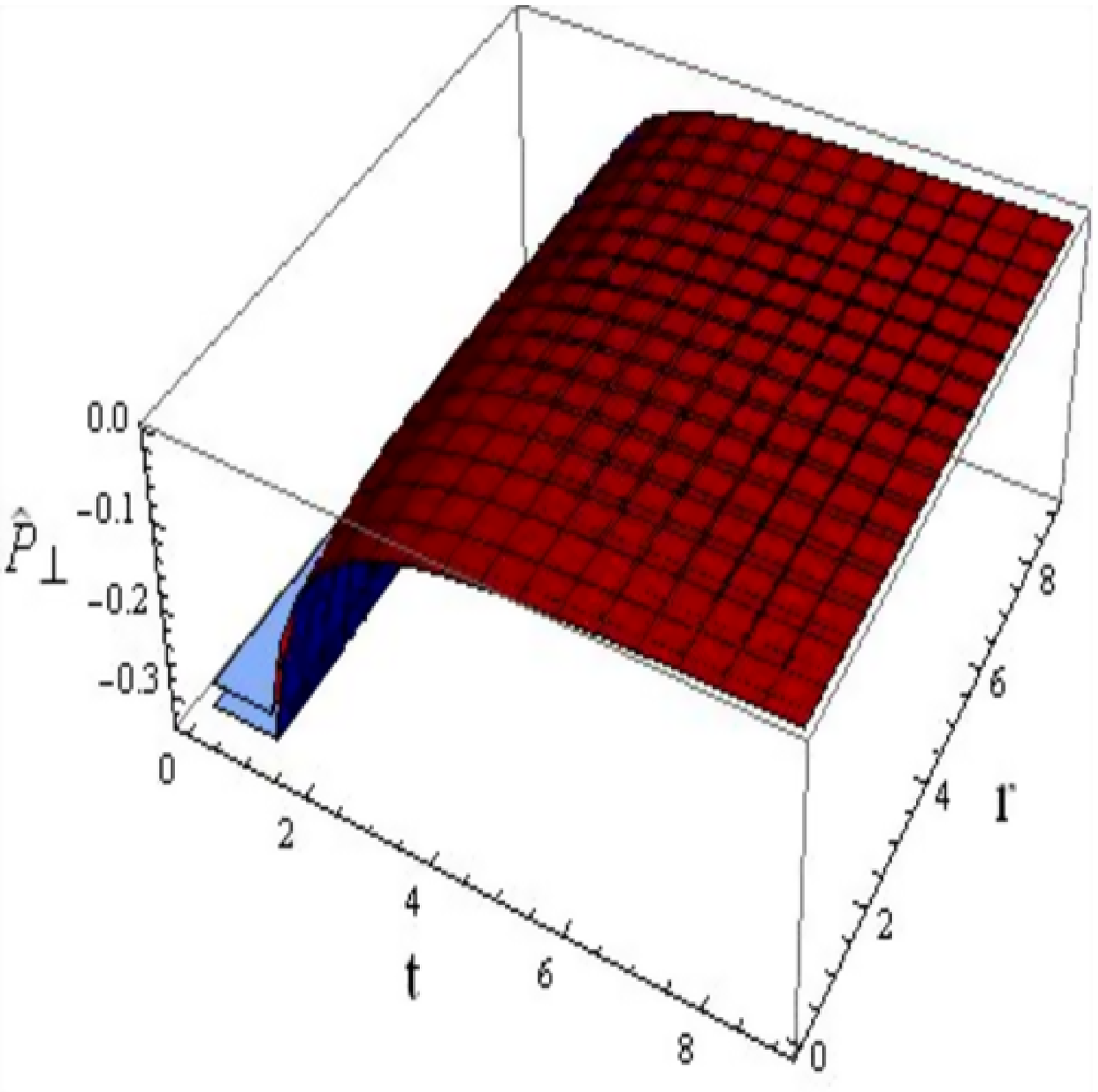,width=0.4\linewidth}\epsfig{file=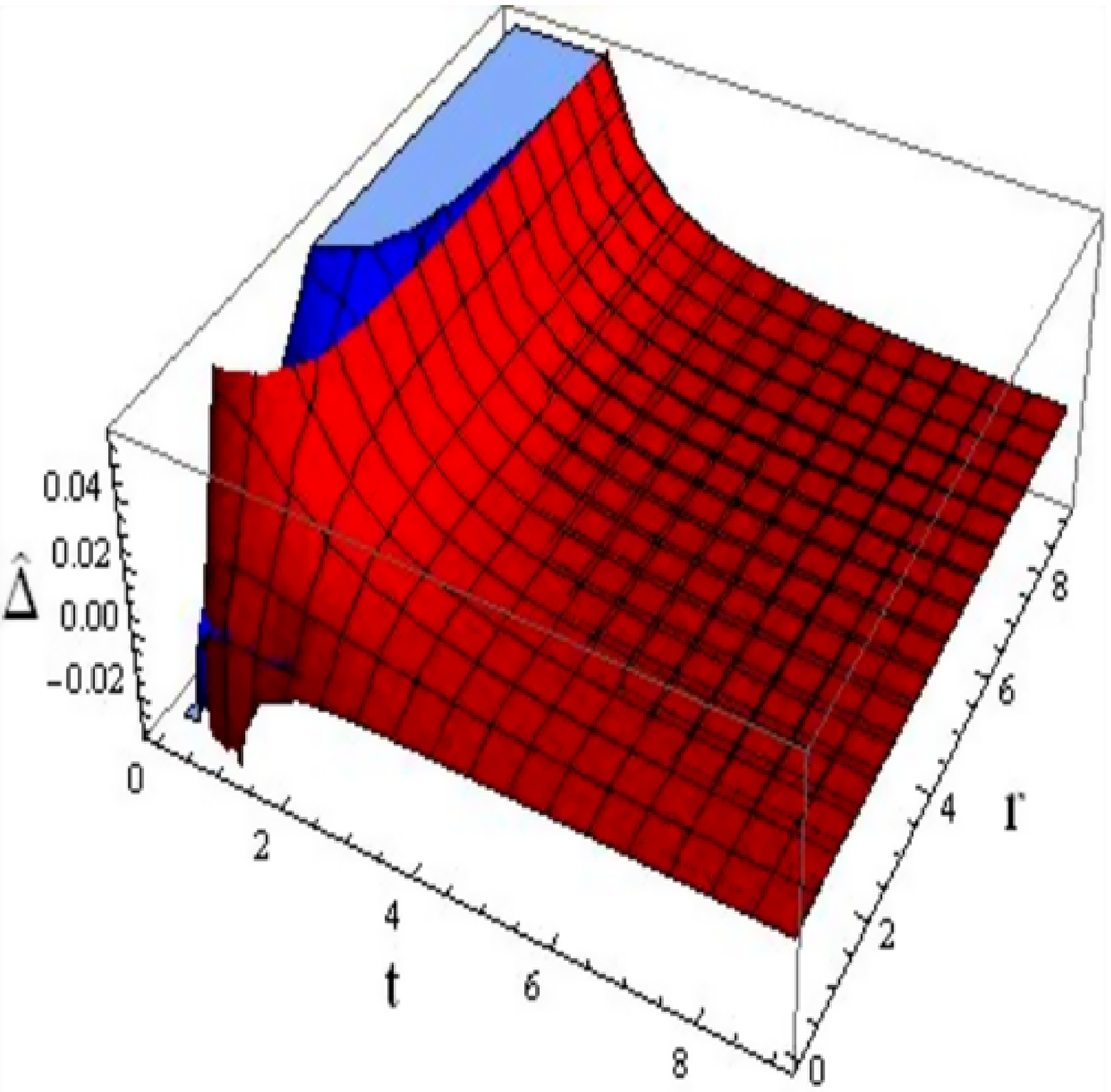,width=0.4\linewidth}
\caption{Plots of matter variables and anisotropy with $\omega=-1$
for $\delta=0.1$ (blue) and $0.9$ (red).}
\end{figure}
\begin{figure}\center
\epsfig{file=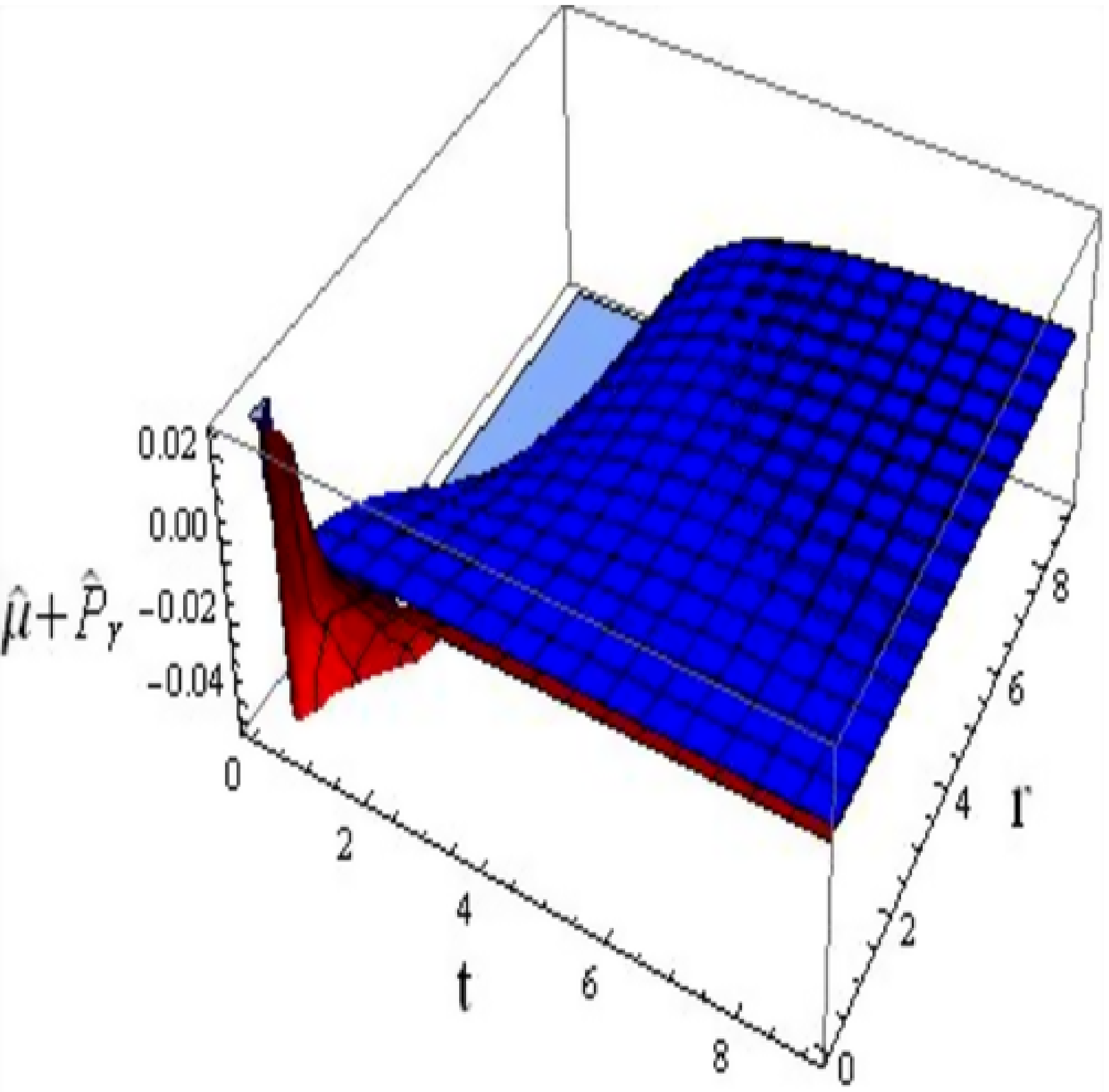,width=0.4\linewidth}\epsfig{file=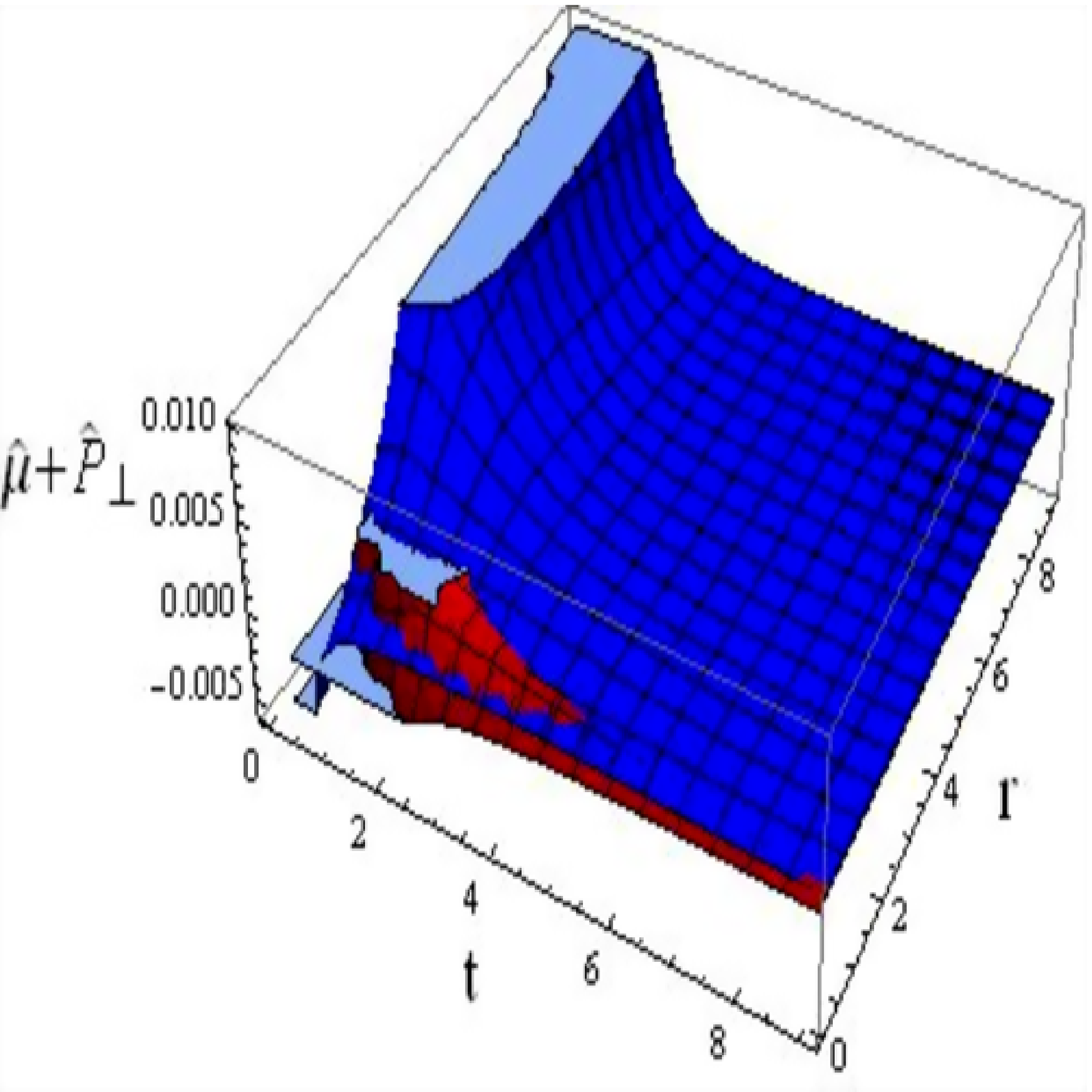,width=0.4\linewidth}
\epsfig{file=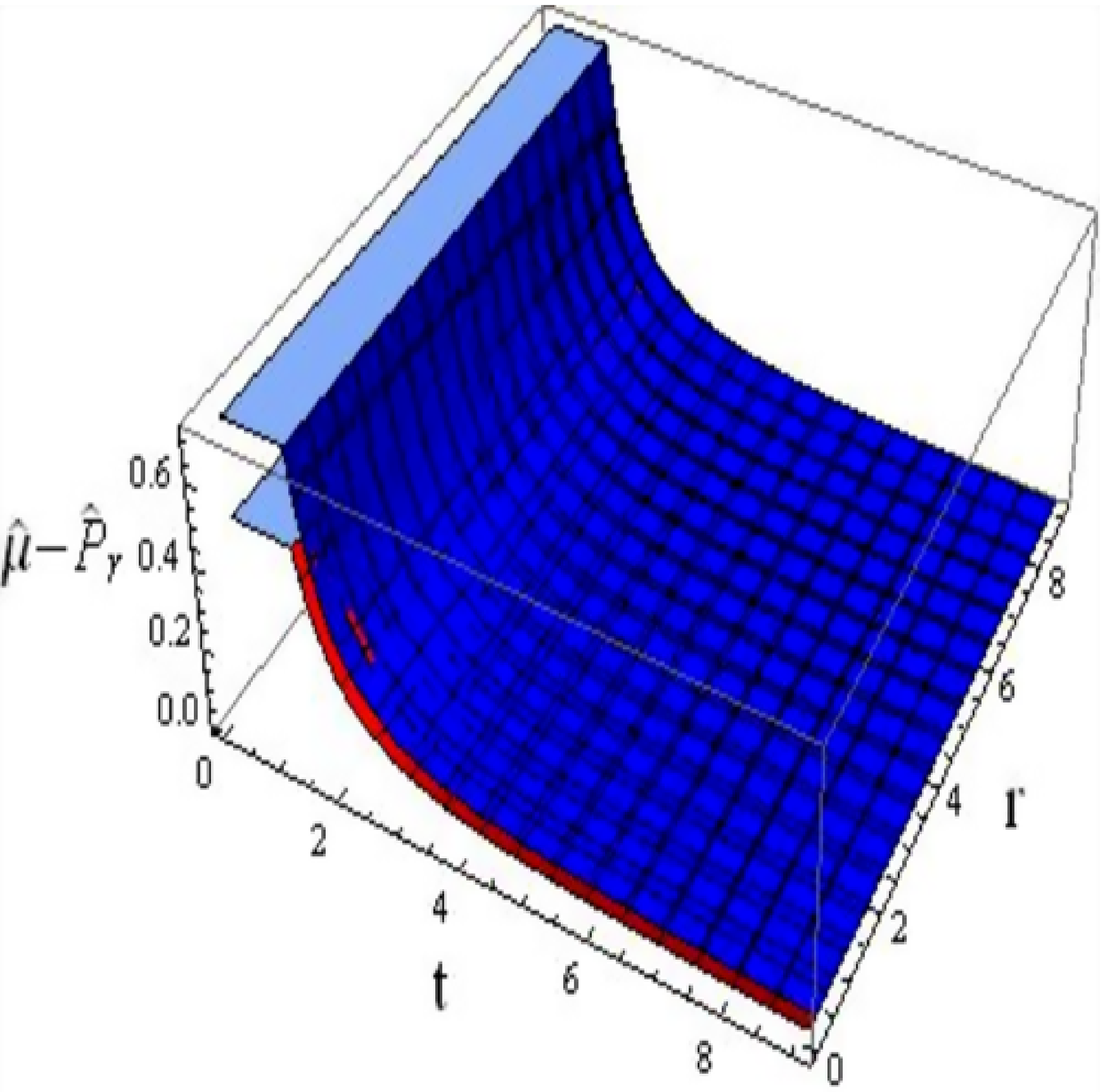,width=0.4\linewidth}\epsfig{file=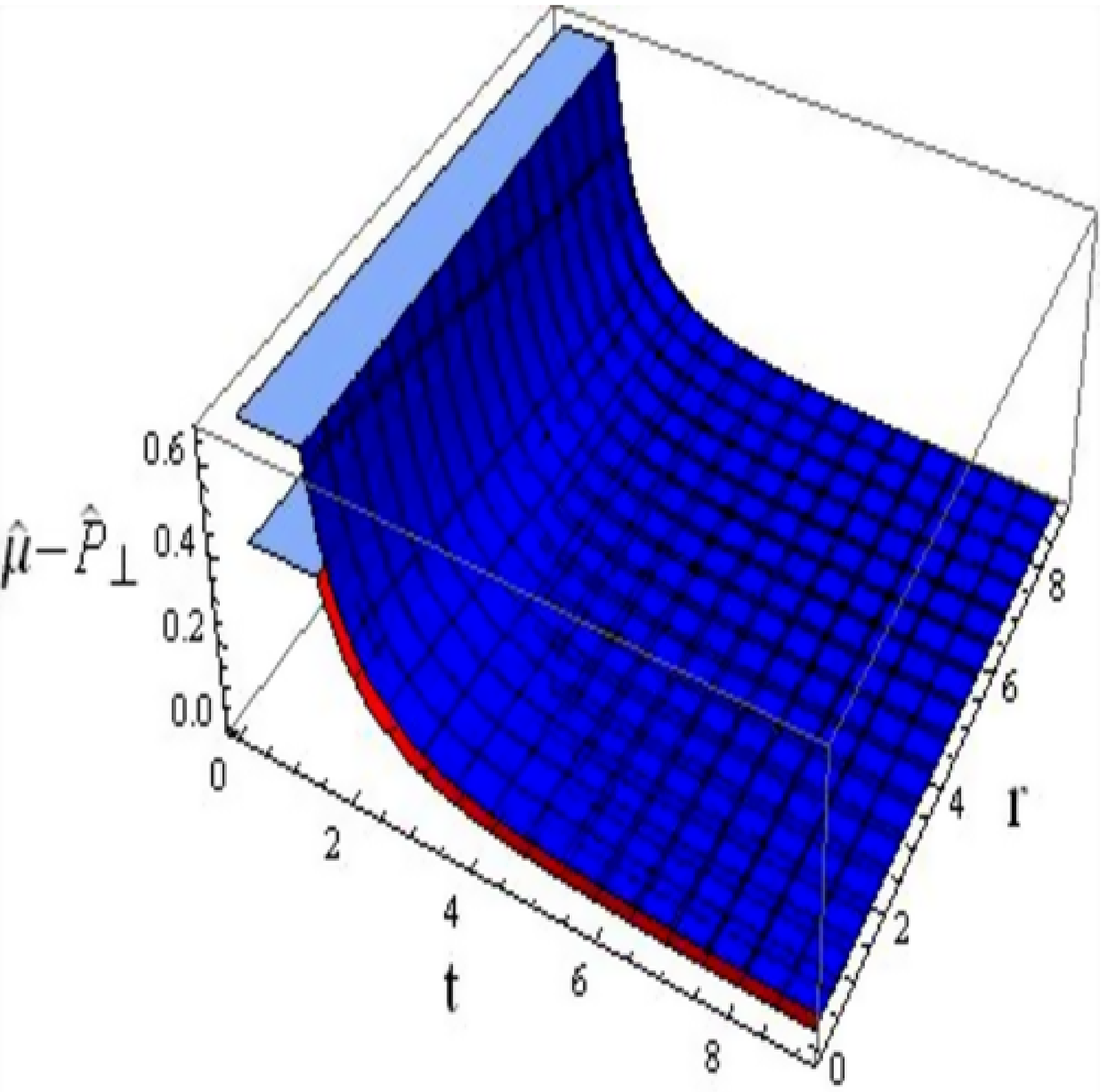,width=0.4\linewidth}
\epsfig{file=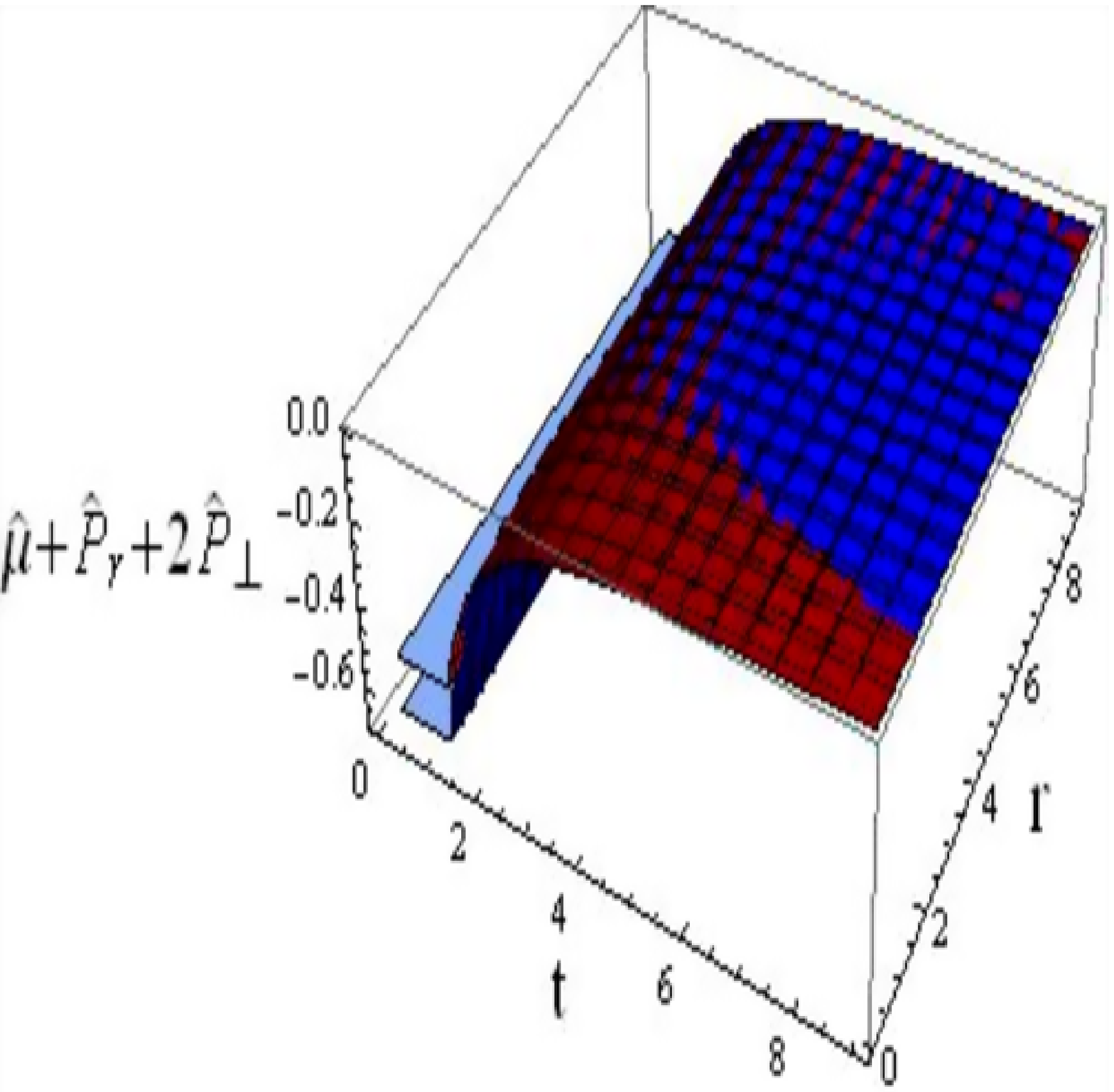,width=0.4\linewidth} \caption{Plots of
$\mathbb{EC}s$ with $\omega=-1$ for $\delta=0.1$ (blue) and $0.9$
(red).}
\end{figure}
\begin{figure}\center
\epsfig{file=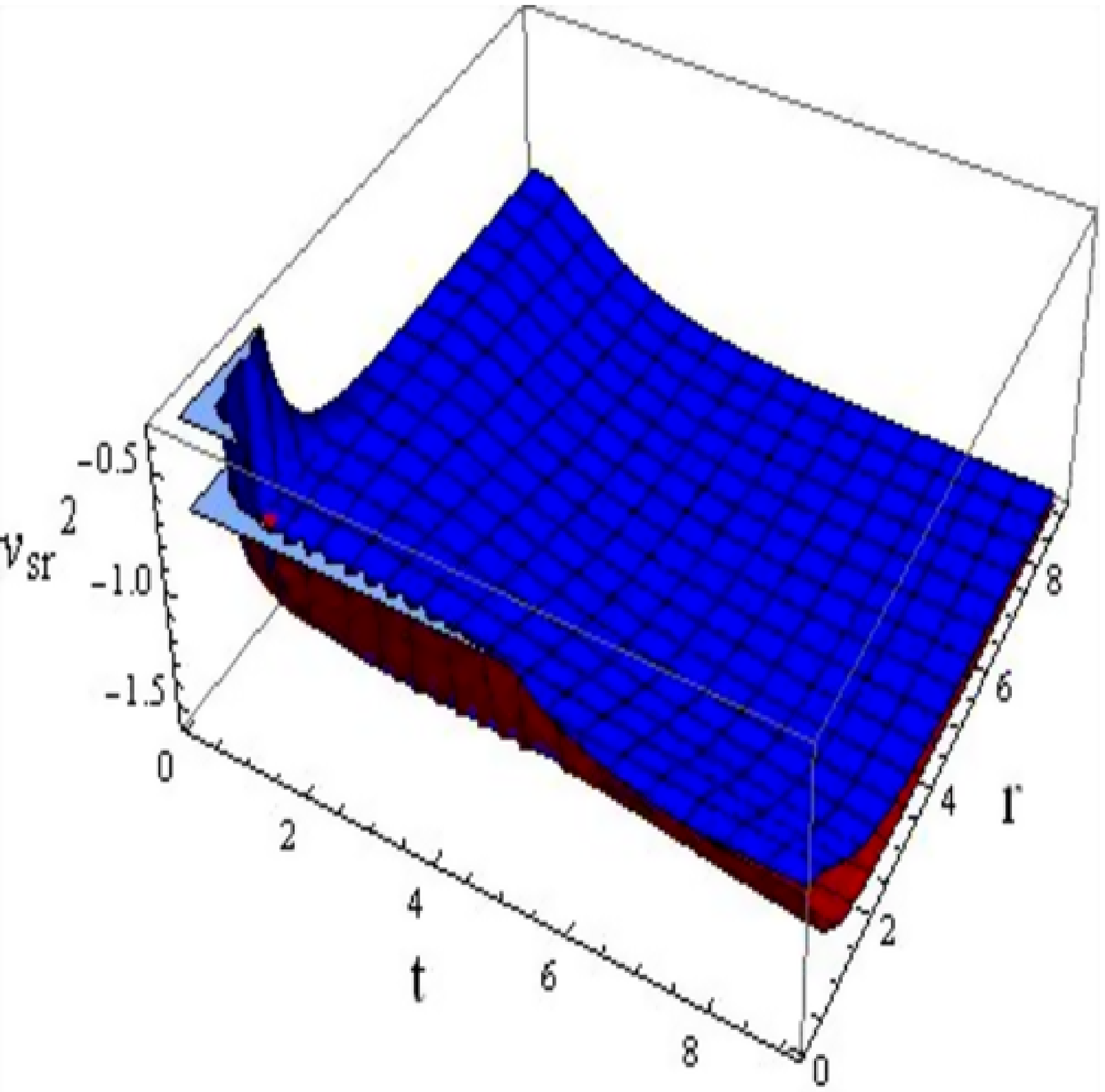,width=0.4\linewidth}\epsfig{file=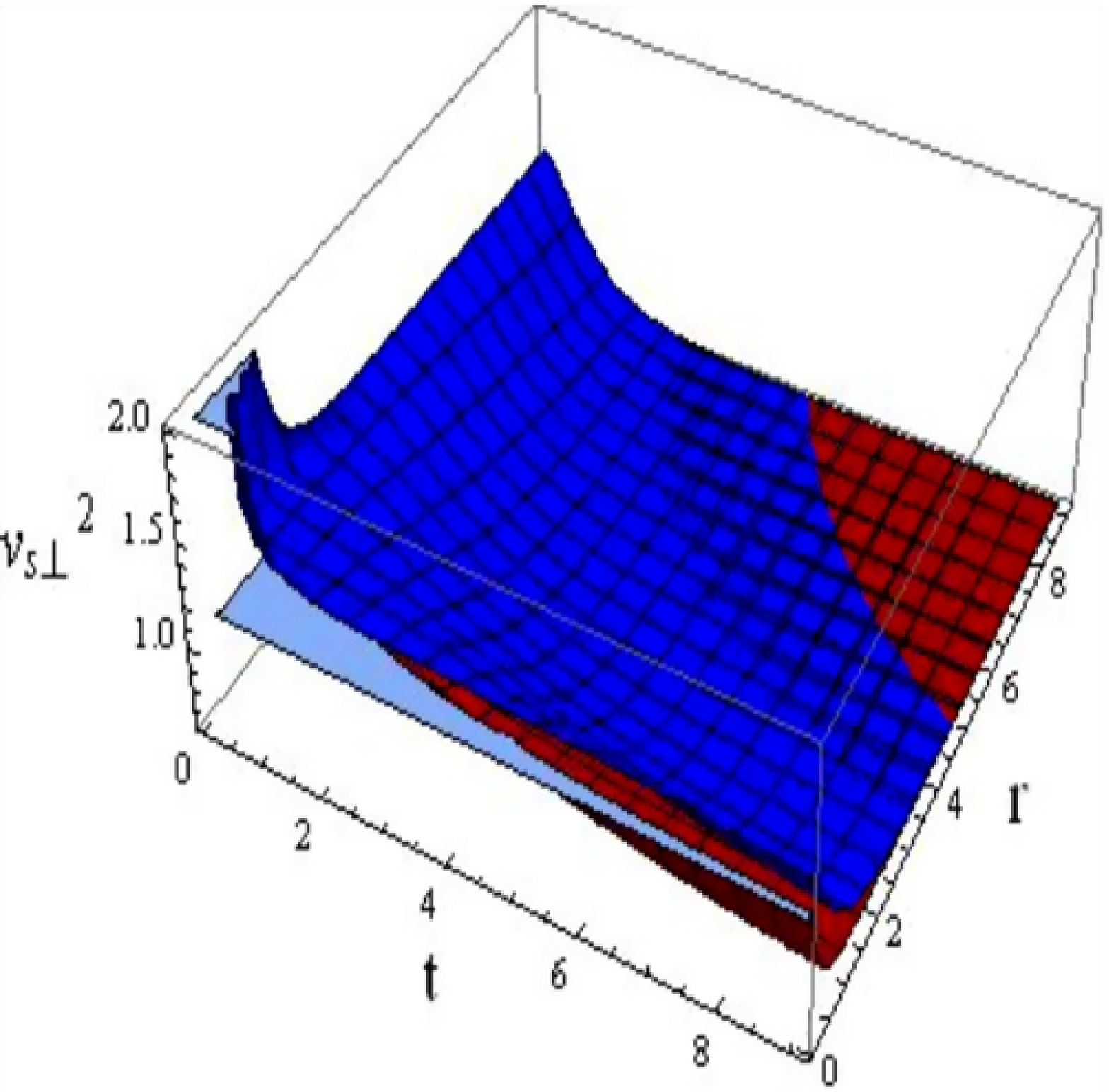,width=0.4\linewidth}
\epsfig{file=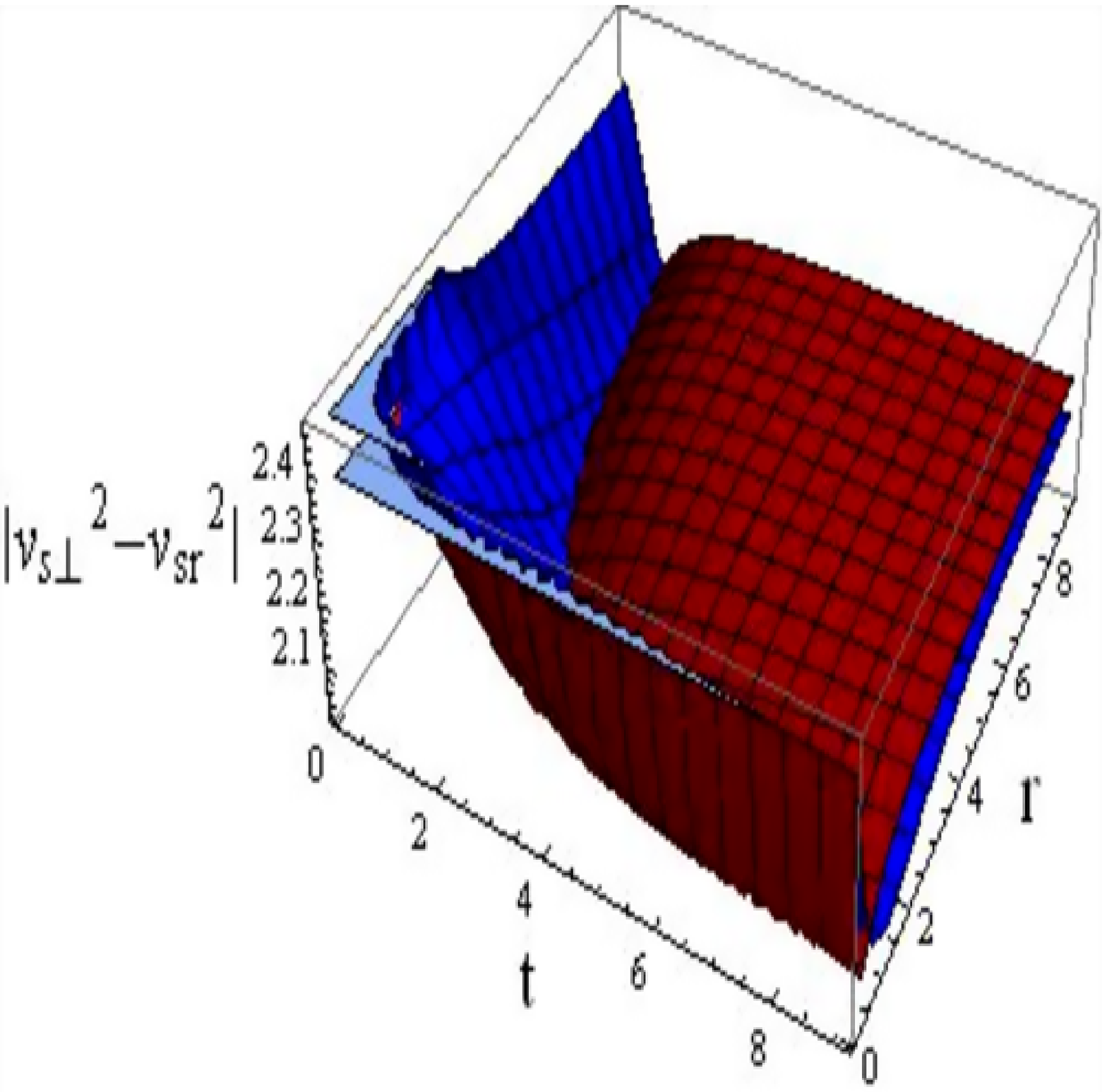,width=0.4\linewidth}\epsfig{file=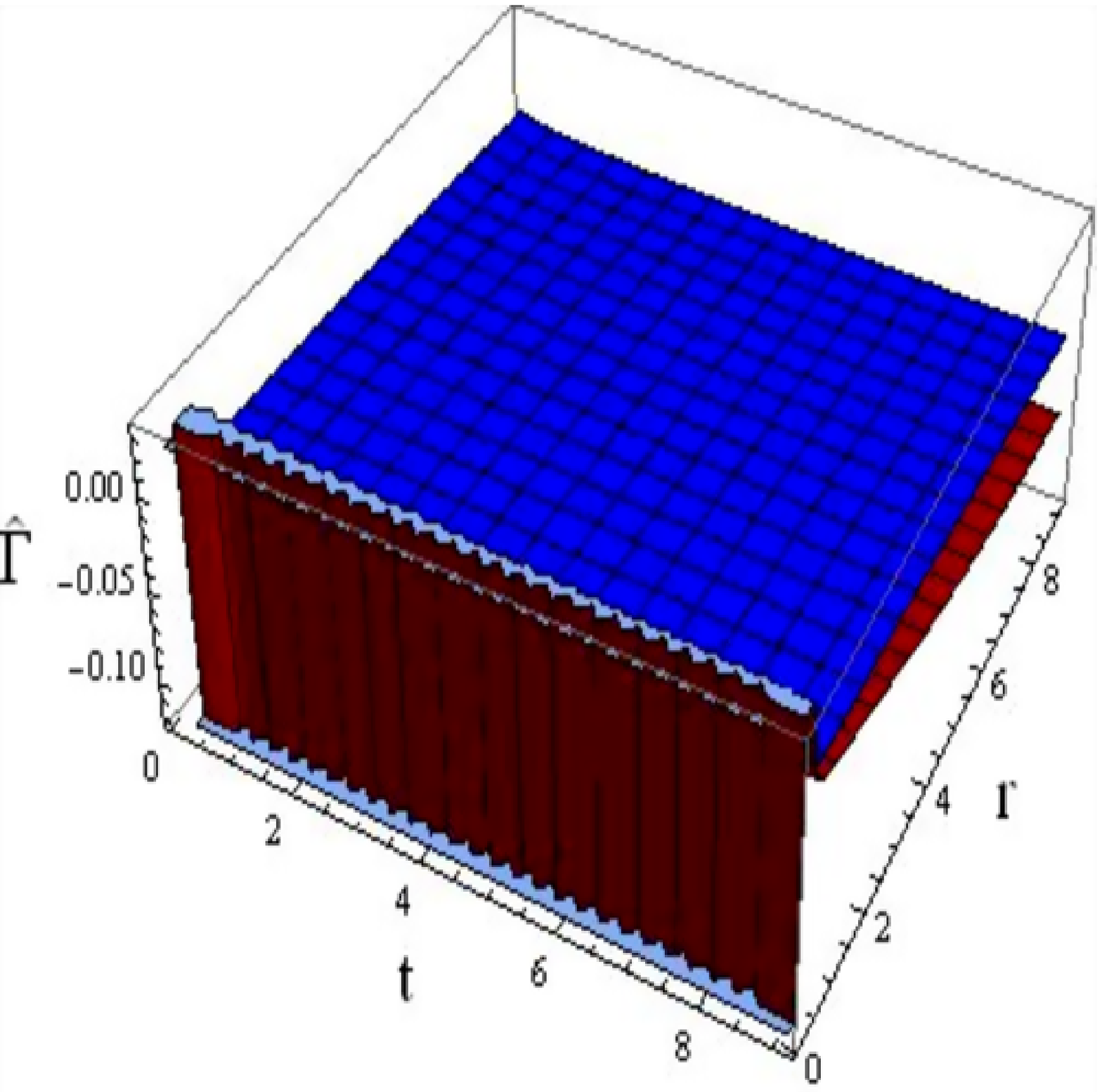,width=0.4\linewidth}
\caption{Plots of $v^{2}_{sr}$, $v^{2}_{s\bot}$,
$|v^{2}_{s\bot}-v^{2}_{sr}|$ and adiabatic index with $\omega=-1$
for $\delta=0.1$ (blue) and $0.9$ (red).}
\end{figure}

\section{Conclusions}

In this paper, we formulate an anisotropic extension of a non-static
spherically symmetric system with the help of MGD in the framework
of $f(\mathbb{R},\mathbb{T})$ theory. To apply decoupling technique,
we choose a linear model as
$f(\mathbb{R},\mathbb{T})=\mathbb{R}+2\varpi\mathbb{T}$ that makes
our results meaningful. The pressure anisotropy is induced in the
seed matter distribution by adding an additional source in the
$\mathbb{EMT}$ whose influence can be controlled by the decoupling
parameter $\delta$. The transformation on $g_{rr}$ component divides
the modified field equations into two arrays. The first of them
characterizes an isotropic FLRW spacetime, and we assume the scale
factor in its power-law form to deal with it. A linear
$\mathbb{E}o\mathbb{S}$ combining energy density and pressure
through a parameter $\omega$ has also been taken to discuss some
different cosmic phases. The other set contains anisotropic effects
that has been solved by employing density-like constraint. Finally,
we have obtained anisotropic cosmological model
\eqref{42}-\eqref{42c} and discussed the impact of $\delta$ by
taking its different values. The physical feasibility of this
extended model has also been studied for different values of
$\omega$ (representing distinct cosmic eras) for $\mathrm{k} = 0$.
We briefly illustrate the main results in the following.
\begin{itemize}
\item The energy density and pressure ingredients show positive and decreasing profile with respect to time
in the radiation-dominated era ($\omega=\frac{1}{3}$), which
suggests the expansion phase of our universe (Figure \textbf{1}).
The results are found to be viable and stable for all the considered
values of parameters in this regime (Figures \textbf{2,~3}).
\item In the matter-dominated era ($\omega=0$), the behavior of state determinants have been found consistent
with the previous cosmic model. However, we observe more dense
profile as compared to the radiation-dominated era (Figure
\textbf{4}). We have also observed a viable and stable model
(Figures \textbf{5,~6}).
\item In the vacuum energy dominated phase ($\omega=-1$), the energy density exhibits decreasing trend with time. However,
both the pressure components are negative that indicate the
existence of an unknown force with strong repulsive nature (Figure
\textbf{7}). The $\mathbb{EC}s$ as well as stability criteria have
been violated, thus this epoch is not viable and stable for any
value of $\delta$ (Figures \textbf{8,~9}).
\end{itemize}

It is noteworthy here that the parameter $\delta$ is inversely
proportional to the energy density and directly related with
radial/tangetial pressures in all cosmic periods. We conclude that
$f(\mathbb{R},\mathbb{T})$ gravity offers more efficient results as
compared to Brans-Dicke gravity, as the matter-dominated era was
found to be unstable in that case \cite{16a}. All our outcomes
ultimately reduce to $\mathbb{GR}$ for $\varpi=0$.

\end{document}